\documentclass[aps,prb,reprint,showpacs,groupedaddress,
footinbib,citeautoscript]{revtex4-1}

\usepackage{graphicx}

\usepackage[utf8]{inputenc}
\usepackage{csquotes}
\usepackage[american]{babel}
\usepackage[T1]{fontenc}
\usepackage{enumerate}
\usepackage{mdwlist}
\usepackage[activate=normal]{pdfcprot}
\usepackage{bbding}
\usepackage{color}
\usepackage{amssymb}
\usepackage{amsmath}
\usepackage{amsfonts}
\usepackage{mathrsfs}

\usepackage{bm}
\usepackage{dcolumn}
\usepackage{color}
\usepackage[colorlinks=true,citecolor=blue]{hyperref}
\hypersetup{colorlinks=true,citecolor=blue,linkcolor=red
,urlcolor=blue}

\graphicspath{{figures/}}

\begin{document}
\title{Supercurrent reversal in Josephson junctions based on bilayer graphene flakes}

\author{Babak Zare Rameshti}
\email{b.zare@iasbs.ac.ir}
\author{Malek Zareyan}
\thanks{Deceased 24 Feb 2014.}
\author{Ali G. Moghaddam}
\email{agorbanz@iasbs.ac.ir}
\affiliation{Department of Physics, Institute for Advanced Studies
in Basic Sciences (IASBS), Zanjan 45137-66731, Iran}

\begin{abstract}
We investigate the Josephson effect in a bilayer graphene flake contacted by two monolayer sheet deposited by superconducting electrodes. It is found that when the electrodes are attached to the different layers of the bilayer, the Josephson current is in a $\pi$ state when the bilayer region is undoped and in the absence of vertical bias. Applying doping or bias to the junction reveals $\pi-0$ transitions which can be controlled by varying the temperature and the junction length. The supercurrent reversal here is very different from  the ferromagnetic Josephson junctions where the spin degree of freedom plays the key role. We argue that the scattering processes accompanied by layer and sublattice index change give rise to the scattering phases which their effect varies with doping and the bias. Such scattering phases are responsible for the $\pi-0$ transitions. On the other hand if both of the electrodes are coupled to the same layer of the flake or the flake has AA stacking instead of common AB, the junction will be always in $0$ state since layer or sublattice index is not changed.

\end{abstract}

\maketitle

\section{Introduction}\label{sec:intro}
Starting from a decade ago, two dimensional (2D) atomic layers synthesized and received a huge amount of interest \cite{geim2007}. Graphene, the leading 2D material, has been studied a lot mostly because of promising applications in electronics, chemistry, optics, etc. beside unexpected relativistic-like electronic dispersion arising interest from fundamental point of view~\cite{neto2009}. It has been shown that although the low-energy quasi-particles in single layer graphene are massless Dirac fermions, however, the situation for bilayer graphene (BG) is fundamentally different revealing chiral gapless excitations with quadratic dispersion rather than linear~\cite{novoselov2006,mccann2007,mccann2013}. Subsequently electronic properties of mono- and bilayer graphene are far from each other and as the most famous example due to the so-called Klein tunneling the backscattering is absent in monolayer graphene (MG) while in the case of bilayer, the forward scattering is impossible \cite{katsnelson2006}. Very intriguingly one can simply use a perpendicular electric field to generate a controllable gap in graphene bilayers~\cite{ohta2006,castro2007,wang2009-nat,oostinga2008,mccann2006,min2007}. 
\par
Recently, it has been experimentally revealed that in epitaxial graphene, it is very probable to have step-like monolayer/bilayer (ML/BL) interfaces as well as bilayer flakes connected to monolayer regions~\cite{ji2012,giannazzo2012,clark2014prx}. A natural question arises immediately about the interplay of massless and massive dispersions in the junctions containing both mono- and bilayer graphene regions. Earlier theoretical investigations had proven that in the ML/BL interfaces the transmission probabilities can show a valley dependent asymmetry which suggests their usage in the generation of valley polarized electron beams~\cite{nilsson2007,ando2010,ando2011}. 
Moreover in the presence of perpendicular magnetic fields the emergence of Landau levels with peculiar transport properties has been studied~\cite{koshino2010,tian2013,puls2009,yan2015}. In particular, a rich Landau spectrum has been predicted~\cite{koshino2010} and an asymmetry in the dependence of transport features to the sign of magnetic field and charge carriers has been experimentally observed~\cite{tian2013}. Other theory works have focused on the edge states properties and quantum transport via channels introduced by the interface~\cite{kohmoto2012,hu2012,li2011,yin2013,
wang2014,berahman2014,dragoman2013}. On the other hand, theoretical investigations of the transport through bilayer flake sandwiched between two monolayer nanoribbons have revealed that the conductance oscillates between maximum and zero as a function of bilayer flake length~\cite{brey2010}. 
\begin{figure}[tp]
\includegraphics[width=0.85\linewidth]{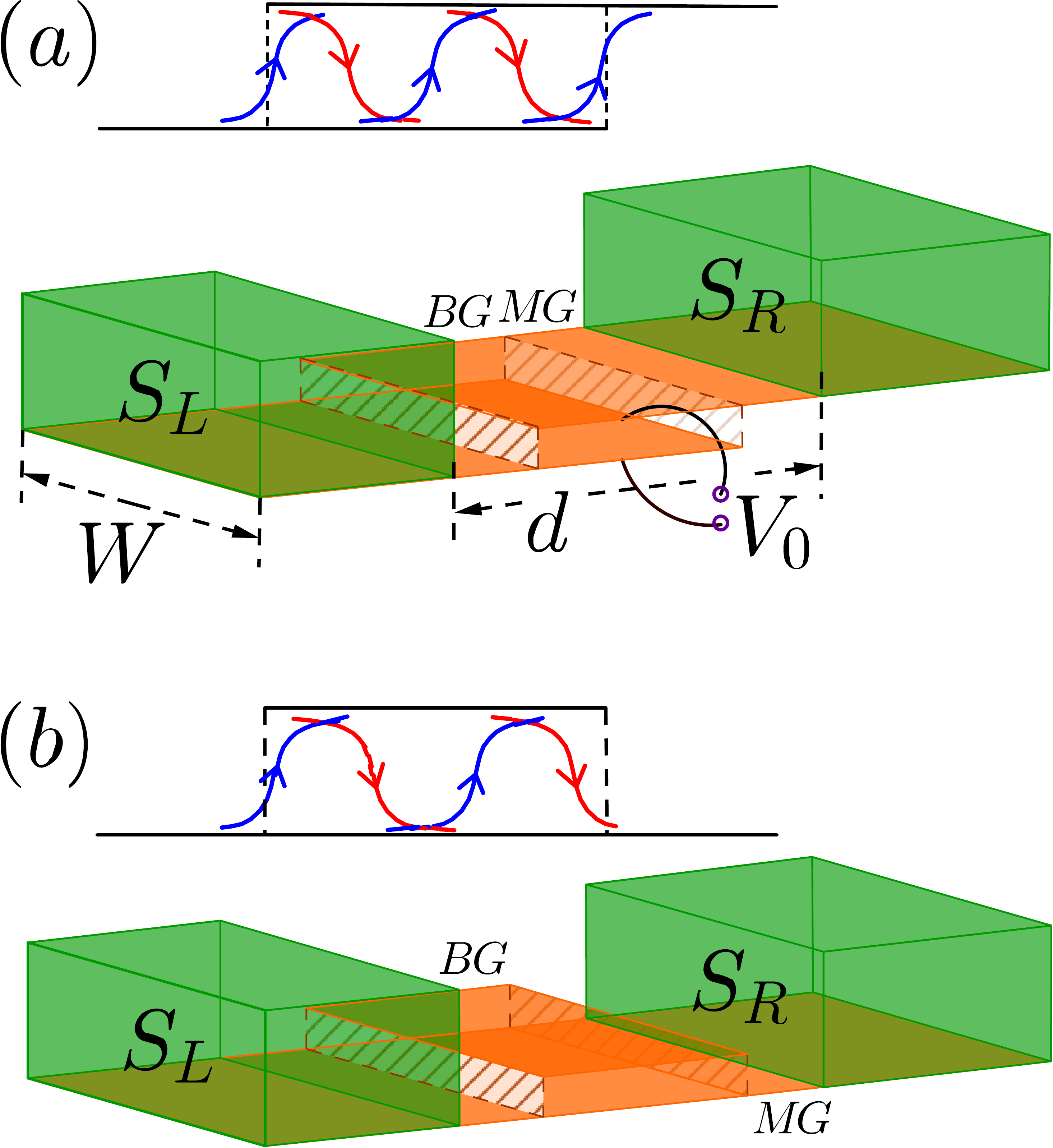}
\caption{
(Color online) Josephson couplings: supercurrent streams through the biased BG flakes embedded between two Josephson coupled superconducting MG leads, $S_{L}$ and $S_{R}$. The length and width of the junction is indicated by $d$ and $W$, respectively. A voltage difference $V_{0}$ is applied perpendicularly between the BG layers. The superconductor either are attached to top and bottom layer of the BG flake (a) or they are connected to the same layer of the BG flake (b). Qualitative description in top of each setups demonstrates the transmission of the quasiparticles through the BG flakes by the red and blue curved arrows.}
\label{fig1}
\end{figure}  
\par
Pioneering work of Beenakker~\cite{beenakker2006,beenakker2008} and successive experimental realization of superconducting proximity effect and supercurrent passing through graphene \cite{heersche} proved that graphene-based superconducting heterostructures can have very interesting properties \cite{titov2006,moghaddam2006,fazio2009,
linder2007,linder2009,
doniach2008,yeyati2010} which some of them are experimentally realized~\cite{bouchiat2009,girit2009,dirks2011,jeong2011,lee2011,lee2015}. Among the variety of theoretical predictions, the so-called specular Andreev reflection, despite several proposals, still waits to be detected \cite{xing2008,bai2008,sun2009,recher2012}. In addition,
when the normal graphene region is replaced with a magnetic one, the well known $0-\pi$ transitions have been predicted to take place with some features different from conventional superconductor-ferromagnet-superconductor (SFS) junctions~\cite{linder2008,moghaddam2008,asano2008}. Intriguingly it has been suggested that one can see the transitions by varying the doping of the magnetic region sandwiched between superconductors \cite{moghaddam2009,zare2014}.
\par
In this paper we investigate the Josephson effect through a BG flake embedded between two monolayer sheets as can be seen in Fig. \ref{fig1}. We find that the first setup in which the two ML leads are coupled to the lower and upper layers of the flake, is in the Josephson $\pi$ state when the doping $\mu$ and vertical bias $V_0$ are zero. Applying gate voltages which results in doping and bias to the flake the device can undergo a $\pi$ to $0$ transitions. Moreover at the presence of finite $\mu$ or $V_0$, the $0-\pi$ transitions occur with temperature and the junction length as well. It must be noted that the reason of such transitions are in contrast with those taking place in the Josephson devices with ferromagnetic weak links. Here the origin of $\pi$ state is the phase factors of transmission coefficients through the flake which influence the superconducting phase dependence of the current carrying Andreev bound states. The appearance of phases is somehow related to the fact that quasiparticles passing through the flake in the setup (a) need to change their layer and sublattice index while in the setup (b) for all the modes the scattering phases are zero.
Varying the bias voltage and doping of the flake changes the scattering properties of the flake and at certain points the overall effect of the scattering phases are diminished leading to the $0$ state Josephson effect.
\par
This article is organized as follows. In Sec.~\ref{sec:model}, we introduce our model and explain the method we use to calculate the supercurrent that streams between two Josephson coupled superconducting leads. In Sec.~\ref{sec:results}, we present and discuss our numerical results that shows the effect due to bilayer step junction as well as bias voltage on supercurrent. Finally, in Sec.~\ref{sec:concl}, we conclude and summarize our main findings.

\section{Model and Basic Formalism}\label{sec:model}
We investigate the supercurrent flowing between two superconducting reservoirs on top of MG regions which are connected through a weak link containing a biased BG flake. 
In fact, there are two different configurations for the setup depending on how the BG flake is connected to the MG regions in the left and right. The left and right reservoirs can be attached to the top and bottom layers of the bilayer, respectively as Fig. \ref{fig1}(a) or both of the monolayer reservoirs are connected to the same layer of the flake as Fig. \ref{fig1}(b). To be precise, we will assume the the MG regions are fully covered by superconductors and only the middle BG flake is in the normal state. We further consider an $s$-wave order parameter, $\Delta({\bf r})=\Delta_{0}e^{i\phi_{L,R}}$ in the left and right superconducting regions and $\Delta({\bf r}) = 0$ inside the flake, where $\Delta_{0}$ is an isotropic gap. 
\par
Similar to any other weak link, the supercurrent across the junction is mainly carried by the discrete bound states which are the result of closed loop motion of the quasiparticles with subgap energies between two superconductors, known as the Andreev bound states (ABS). One can see that it is usually sufficient to find ABS energies in order to calculate the supercurrent carried by them.
In the short junction limit, where the length of the junction ($d$) is much smaller than the coherence length of the superconductor ($\xi$), the current establishes due to the phase dependence of ABS energies and the continuous spectrum does not contribute since its density of states is almost phase independent. The relation between the Josephson current $I$ passing through the junction with the transverse width $W$ and subgap quasiparticles ($\varepsilon_{n}$) at finite temperature $T$ is given by,
\begin{eqnarray}
I=-\frac{2eW}{\hbar}\int dk_{y}\sum_{n}\tanh\left[\frac{\varepsilon_{n}(k_{y})}{2k_{B}T}\right]\frac{d\varepsilon_{n}(k_{y})}{d\phi}\label{eq1}
\end{eqnarray}
The sum is over all ABS energies which are positive  corresponding to the different transverse momenta $k_{y}$. 
\par 
The ABS energies can be calculated in the framework of Bogoliubov-de Gennes (BdG) equations which describe the superconducting correlations between particles and their time-reversed counterparts (holes). Exploiting the BdG equations, the ABS energies will be obtained as the roots of a characteristic equation containing the whole scattering matrix of the junction. This method is based on the fact that the ABS can be assumed as the states scattered completely to themselves inside the junction. The whole scattering matrix consists of two parts: ${\rm S}_{B}$ which describes the normal scatterings of electrons and holes within the bilayer flake and ${\rm S}_{A}$ responsible for the scattering away from the normal-superconducting (NS) interfaces.
Since for subgap energies $\varepsilon<\Delta_{0}$ there are no propagating modes in the superconducting regions $L$ and $R$ then the scattering matrix S$_{A}$ can be defined with
relation $a_{B}=\mathrm{S}_{A}b_{B}$
which relates the outgoing to incoming quasiparticles. This matrix consists of four blocks in electron-hole (Nambu) space corresponding to different processes as,
\begin{eqnarray}
\mathrm{S}_{A}=\begin{pmatrix}
\mathrm{\check S}_{ee} & \mathrm{\check S}_{eh}\\
\mathrm{\check S}_{he} & \mathrm{\check S}_{hh}
\end{pmatrix}\label{eq2}.
\end{eqnarray}
In general each block of the scattering matrix in Nambu space denoted by $\mathrm{\check S}$ have the following form,
\begin{equation}
\mathrm{\check S}=\begin{pmatrix}
\hat{r}_{LL} & \hat{t}_{LR}\\
\hat{t}_{RL} & \hat{r}_{RR}
\end{pmatrix},
\end{equation}
consisting of reflection $\hat{r}$ and transmission $\hat{t}$ matrices which here are $2\times2$ matrices in the space of two layers. The scattering matrix of the NS interfaces involves only the reflection processes and subsequently the corresponding normal and Andreev reflection parts are block-diagonal as below,
\begin{eqnarray}
\mathrm{\check S}_{ee}
=\begin{pmatrix}
\mathrm{\hat S}^L_{ee} & 0\\
0 & \mathrm{\hat S}^R_{ee}
\end{pmatrix}~~,~~~ \mathrm{\check  S}_{eh}=\begin{pmatrix}
\mathrm{\hat S}^L_{eh}  & 0\\
0 & \mathrm{\hat S}^R_{eh}
\end{pmatrix}.
\end{eqnarray}
Now we should remind that at the left and right NS interfaces, depending on the configuration, either different layers (setup a in Fig. \ref{fig1}), or only one of the layers (setup b in Fig. \ref{fig1})
are involved in the reflection processes. As a result the matrices for the two setups are obtained respectively as below,  
\begin{eqnarray}
(a):~&&\nonumber\\
&&\mathrm{\hat S}^L_{ee}=-\alpha(\varepsilon)\hat{\tau}_{u}~~,~~\mathrm{\hat S}^L_{eh}=\beta(\varepsilon)e^{i\phi/2}
\hat{\tau}_{u},~~~~~\nonumber\\
&&\mathrm{\hat S}^R_{ee}=\alpha(\varepsilon)\hat{\tau}_{d}
~~,~~\mathrm{\hat S}^R_{eh}=\beta(\varepsilon)e^{-i\phi/2}\hat{\tau}_{d},\\
(b):~&&\nonumber\\
&&\mathrm{\hat S}^L_{ee}=-\alpha(\varepsilon)\hat{\tau}_{u}
~~,~~\mathrm{\hat S}^L_{eh}=\beta(\varepsilon)e^{i\phi/2}
\hat{\tau}_{u},\nonumber\\
&&\mathrm{\hat S}^R_{ee}=\alpha(\varepsilon)\hat{\tau}_{u}
~~,~~\mathrm{\hat S}^R_{eh}=\beta(\varepsilon)e^{-i\phi/2}\hat{\tau}_{u},
\end{eqnarray}
with $\hat{\tau}_{u, v}
=(\hat{\tau}_{0}\pm\hat{\tau}_{z})/2$ in which $\hat{\tau}_{i}$ being the Pauli matrices in the layer space and $\mathrm{\check S}_{hh}=-\mathrm{\check S}_{ee}$, $\mathrm{\hat S}^L_{he}={\rm exp}(-i\phi)\mathrm{\hat S}^L_{eh}$, $\mathrm{\hat S}^R_{he}={\rm exp}(i\phi)\mathrm{\hat S}^R_{eh}$.
The normal and Andreev reflection amplitudes are  obtained for the MG-based NS interfaces in Ref. \onlinecite{beenakker2006}
as $\alpha(\varepsilon)=i\zeta\sin\theta/
[(\varepsilon/\Delta_{0})
\cos\theta+\zeta]$ and $\beta(\varepsilon)=\cos\theta/[(\varepsilon/\Delta_{0})\cos\theta+\zeta]$  
in which $\zeta=\sqrt{(\varepsilon/
\Delta_{0})^{2}-1}$ and $\theta$ is the incidence angle.
\par
The scattering matrix of the normal BG region S$_{B}$ relates two coefficient vectors of transmitted and reflected as $b_{B}=\mathrm{S}_{B}a_{B}$. Owing to the fact that matrix S$_{B}$ does not couple the electrons and holes together, it has a block-diagonal form and given by,
\begin{eqnarray}
\mathrm{S}_{B}=\begin{pmatrix}
\mathrm{\hat S}(\varepsilon) & 0\\
0 & \mathrm{\hat S}^{\ast}(-\varepsilon)
\end{pmatrix},\label{eq3}
\end{eqnarray}
Here S$(\varepsilon)$ and $\mathrm{S}^{\ast}(-\varepsilon)$ are the unitary and symmetric scattering matrices governing the scattering properties of the electrons and holes. The reflection and transmission matrices $\hat{r}$ and $\hat{t}$ for the BG flake embedded between two MG are given by,
\begin{eqnarray}
(a):~&&\nonumber\\
&&\hat{r}_{LL}=r_{11}\hat{\tau}_{u}~~,~~~\hat{r}_{RR}=r^{'}_{22}\hat{\tau}_{d}
~~~~~\nonumber\\
&&\hat{t}_{LR}=t^{'}_{12}\hat{\tau}_{+}~~,~~~
\hat{t}_{RL}=t_{21}\hat{\tau}_{-}\\
(b):~&&\nonumber\\
&&\hat{r}_{LL}=r_{11}\hat{\tau}_{u}~~,~~~\hat{r}_{RR}=r^{'}_{11}\hat{\tau}_{u}
\nonumber\\
&&\hat{t}_{LR}=t^{'}_{11}\hat{\tau}_{u}~~,~~~
\hat{t}_{RL}=t_{11}\hat{\tau}_{u}\label{eq4}
\end{eqnarray}
for the two different setups, respectively, with $\hat{\tau}_{\pm}=(\hat{\tau}_x\pm i\hat{\tau}_y)/2$. The labels $1$ and $2$ denote the two layers of the BG flake.
Therefore it is clear from above relations that in the setup (b) the layer index is conserved, however, in the setup (a) since the only way to pass through the scattering region is via the interlayer hopping between the two layers it is not conserved anymore.
Transmission $t_{ij}$ and reflection $r_{ij}$ amplitudes can be obtained by matching the wavefunctions at the ML/BL boundaries. In order to complete the construction of the layer resolved scattering matrix, the excitation spectrum of BG as well as MG and their wavefunctions as the scattering basis are needed.
\par
The low-energy Hamiltonian of BG flake in the presence of layer asymmetry due to the bias voltage applied perpendicularly and in the vicinity of non-equivalent corners of Brillouin zone $K$ and $K'$ is given by,
\begin{eqnarray}
\mathcal{H}_{\mathrm{BG}}=\begin{pmatrix}
-\mu+\frac{V}{2} & v_{F}p_{-} & t_{\perp} & 0\\
v_{F}p_{+} & -\mu+\frac{V}{2} & 0 & 0\\
t_{\perp} & 0 & -\mu-\frac{V}{2} & v_{F}p_{+}\\
0 & 0 & v_{F}p_{-} & -\mu-\frac{V}{2}
\end{pmatrix},~~~~\label{BG-H}
\end{eqnarray}
in the layer and sub-lattice spaces
with eigenfunctions of the form $\Phi^{\dagger}_{{\bf p}}=(c_{A1,{\bf p}}, c_{B1,{\bf p}}, c_{B2,{\bf p}}, c_{A2,{\bf p}})$.
Here $p_{\pm}=p_{x}\pm ip_{y}$ and ${\bf p}=(p_{x}, p_{y})$ is the two dimensional momentum measured relative to the $K$ point, $v_{F}\approx 10^{6}m/s$ is the Fermi velocity and $\mu$ being the chemical potential. The nearest neighbor atoms in two layers $A1$ and $B2$ are connected by interlayer hopping term $t_{\perp}\simeq 0.3~ev$ which tends to equalize the charge densities in the two layers. The potential difference between the two layers is involved by the parameter $V$ which opens a gap in the spectrum in contrast to the case of gapless spectrum $V=0$. The vertical bias through the BG also works against the interlayer hopping since it generates charge imbalance between the two layers. Taking plane wave basis we end up for the excitation spectrum of the Hamiltonian with four energy bands given by,
\begin{eqnarray}
(\varepsilon_{\mathrm{BG}}+\mu)^{2}&=&(v_{F}p)^{2}+(V^{2}/4)+(t^{2}_{\perp}/2)\nonumber\\
&&\qquad\pm\sqrt{(t_{\perp}^{2}/2)^{2}+(v_{F}p)^{2}(V^{2}+t_{\perp}^{2})},~~~~~\label{eq6}
\end{eqnarray}
the corresponding eigenvector reads,
\begin{eqnarray}
\Phi^{\dagger}_{{\bf p}}=\mathcal{A}\begin{pmatrix}
\gamma_{-} & v_{F}p_{+} & \eta & \frac{\eta}{\gamma_{+}}v_{F}p_{-}
\end{pmatrix}.\label{eq7}
\end{eqnarray}
where $\eta=[\gamma_{-}^{2}-(v_{F}p)^{2}]/t_{\perp}$ with $\gamma_{\pm}=\varepsilon_{\mathrm{BG}}+\mu\pm V/2$ and the normalization factor is $\mathcal{A}=\left[4v_{F}p_{x}\left(\gamma_{-}-(\varepsilon_{\mathrm{BG}}+\mu)V\eta/\gamma_{+}\right)\right]^{-1/2}$. 
\par
In order to finds the scattering matrix S$_B$ we need the eigenfunctions inside the left and right MG regions which are immediately coupled to the superconductors. The wave-functions are represented in the space of two sublattices as $\Phi^{\dagger}_{{\bf k}}=(c_{A1, {\bf k}}, c_{B1, {\bf k}})$, and the two dimensional Dirac Hamiltonian around $K$ and $K^{\prime}$ points reads,
\begin{eqnarray}
\mathcal{H}_{\mathrm{MG}}=\begin{pmatrix}
\mu^{\prime} & v_{F}(k_{x}-ik_{y})\\
v_{F}(k_{x}+ik_{y}) & \mu^{\prime}
\end{pmatrix}\label{eq8}
\end{eqnarray}
Here the two-dimensional momentum is ${\bf k}=(k_{x}, k_{y})$ and $\mu^{\prime}$ is the chemical potential in the MG so that the excitation energy given by $\varepsilon_{\mathrm{MG}}=\mu^{\prime}\pm v_{F}k$ is measured with respect to the Fermi energy in MG. The corresponding eigenvector are,
\begin{eqnarray}
\Phi_{{\bf k},\xi}^{\dagger}=\left(2\cos\theta\right)^{-1/2}\begin{pmatrix}
1 & e^{i\theta}
\end{pmatrix}\label{eq9}
\end{eqnarray}
with $\xi=L,R$ and $\theta=\arctan(k_{y}/k_{x})$. Since the eigenvectors will be utilized as scattering basis they need to be normalized in a way that each state carries the same amount of quasiparticles current density. 
\par
We are now in a position to write down the wavefunctions in the left, middle and right region as a linear superposition of constructed scattering basis Eqs. (\ref{eq7}) and (\ref{eq9}),
\begin{eqnarray}
\Psi_{L}&=&\Phi^{+}_{{\bf k}, L}e^{i(k_{x}x+k_{y}y)}+r\Phi_{{\bf k}, L}^{-}e^{i(-k_{x}x+k_{y}y)}\nonumber\\
\Psi_{\mathrm{B}}&=&a\Phi_{{\bf p}, 1}^{+}e^{i(p_{1x}x+k_{y}y)}+b\Phi^{+}_{{\bf p}, 2}e^{i(p_{2x}x+k_{y}y)}\nonumber\\
&+&c\Phi^{-}_{{\bf p}, 1}e^{i(-p_{1x}x+k_{y}y)}+d\Phi^{-}_{{\bf p}, 2}e^{i(-p_{2x}x+k_{y}y)}\nonumber\\
\Psi_{R}&=&t\Phi_{{\bf k}, R}^{+}e^{i(k_{x}(x-d)+k_{y}y)}\label{eq10}
\end{eqnarray}
here $+~(-)$ denote right (left) moving quasiparticles with $r$ and $t$ being the reflection and transmission amplitudes respectively. Scattering coefficients can be computed by
solving the linear system constructed from the matching boundary conditions at the interfaces and subsequently will be used to find the the transport properties across the junction. Thus we need to provide the appropriate boundary conditions at the boundaries $x=0$ and $x=d$. The MGs can be attached either to the same layer of BG or to the different layers. We assume the zig zag boundary conditions at the ML/BL interfaces. The boundary conditions at the left interface $x=0$ are,
\begin{eqnarray}
&&\Psi_{L}(x=0)\left\vert_{A_{1}}=\Psi_{B}(x=0)\right\vert_{A_{1}}\nonumber\\
&&\Psi_{L}(x=0)\left\vert_{B_{1}}=\Psi_{B}(x=0)\right\vert_{B_{1}}\nonumber\\
&&\left.\Psi_{B}(x=0)\right\vert_{A_{2}}=0\label{eq11}
\end{eqnarray}
and for the setup b they will be similar at the right interface as well. However considering the top layer as the connecting layer between the right lead and the scattering region (setup a), at the right interface $x=d$ we must have,
\begin{eqnarray}
&&\Psi_{R}(x=d)\left\vert_{A_{2}}=\Psi_{B}(x=d)\right\vert_{A_{2}},\nonumber\\
&&\Psi_{R}(x=d)\left\vert_{B_{2}}=\Psi_{B}(x=d)\right\vert_{B_{2}},\nonumber\\
&&\left.\Psi_{B}(x=d)\right\vert_{B_{1}}=0.\label{eq12}
\end{eqnarray}
\par
Having both scattering matrices $S_{\rm A}$ and $S_{\rm B}$, the general condition for bound states $a_{B}=\mathrm{S}_{A}\mathrm{S}_{B}a_{B}$ implies that $
\mathrm{det}(1-\mathrm{S}_{A}\mathrm{S}_{B})=0$.
Subsequently from Eqs. (\ref{eq2}) and (\ref{eq3}), we find the following characteristic equation for the ABS energies,
\begin{eqnarray}
&&{\rm det}\left\{ 1-\mathrm{\check S}_{ee}[\hat{\mathrm{S}}(\varepsilon)-\hat{\mathrm{S}}^{*}(-\varepsilon)]-\mathrm{\check S}_{ee}\hat{\mathrm{S}}(\varepsilon)\mathrm{\check S}_{ee}\hat{\mathrm{S}}^{*}(-\varepsilon)\right.\nonumber\\
&&~~~~\left.-[1-\mathrm{\check S}_{ee}\hat{\mathrm{S}}(\varepsilon)]\mathrm{\check S}_{he}\hat{\mathrm{S}}(\varepsilon)[1-\mathrm{\check S}_{ee}\hat{\mathrm{S}}(\varepsilon)]^{-1}\mathrm{\check S}_{eh}\hat{\mathrm{S}}^{*}(-\varepsilon)\right\}=0,\nonumber\\&&\label{eq13}
\end{eqnarray}
which can be used to find the Josephson current across the junction given by Eq. (\ref{eq1}).

\section{Results and Discussion}\label{sec:results}
\begin{figure}
\includegraphics[width=0.85\linewidth]{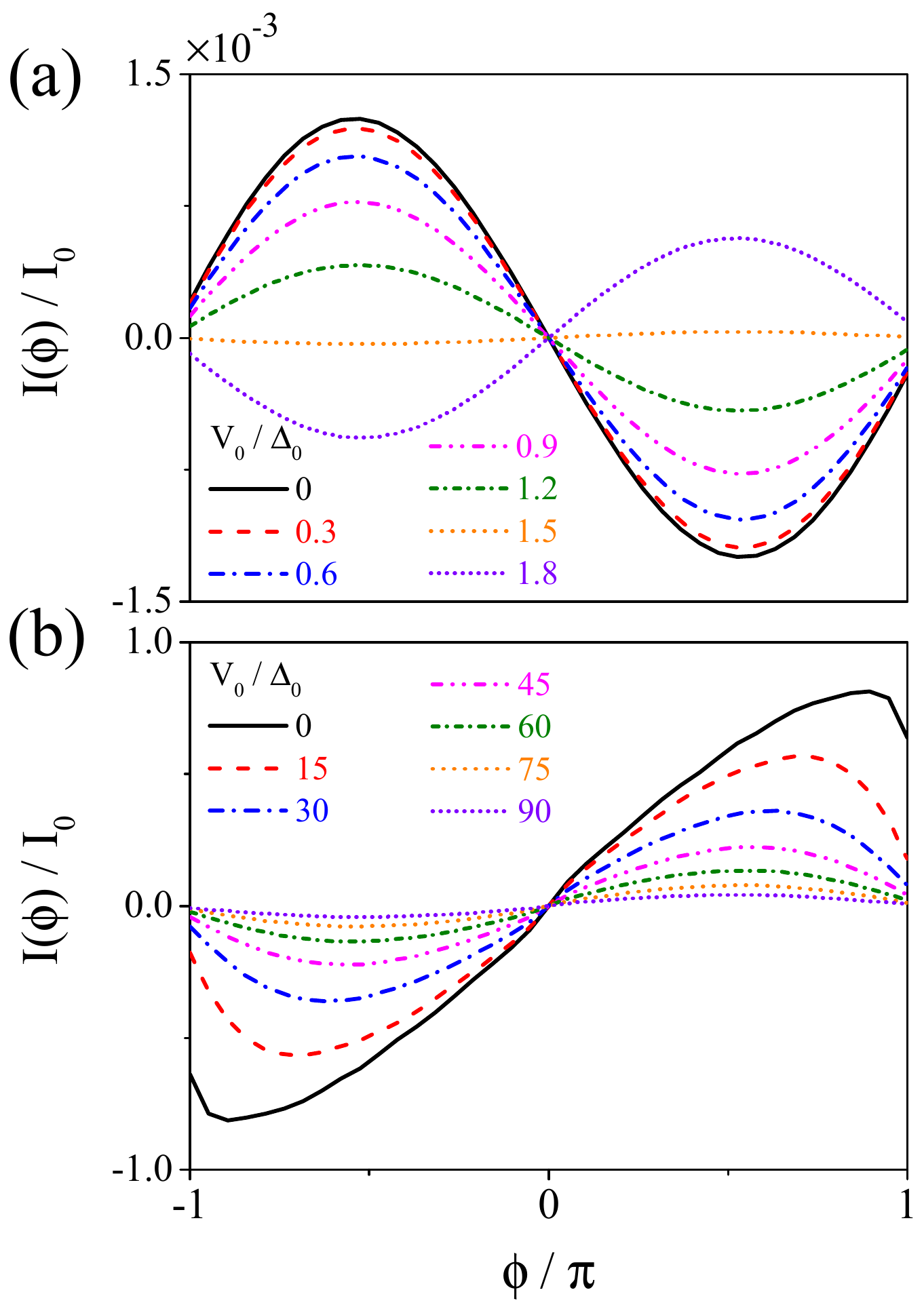}
\caption{
(Color online) The current-phase relation $I(\phi)/I_{0}(T)$ in the short junctions which made of undoped BG flakes are shown at the zero temperature ($T=0$) varying the scaled bias voltage $V_{0}/\Delta_{0}$ for the first setup in (a) and for the second one in (b). The length of the junction is fixed at the $d/l_{\perp}=4$.
}\label{fig2}
\end{figure} 
\par
Here the numerical results for the Josephson current passing through BG flakes between ML regions contacted by superconductors are presented. First we assume the undoped BG with $\mu=0$ and study the current phase relation (CPR) vary the vertical bias. We denote the Fermi wave vector with $k_{F}$ and define $l_{\perp}=\hbar v_{F}/t_{\perp}$ which is a length scale over which
the excitations traveling in the two layers of the BG are coupled. Figure \ref{fig2} shows the supercurrent variations $I(\phi)$ scaled by $I_0(T)=2e\Delta(T)Wk_F/\hbar$ for the two different setups when the junction length is $d/l_{\perp}=4$ and at the zero temperature $T=0$.
We see that at the absence of vertical bias $V_0$ the setup (a) is in the so-called $\pi$-state with $I(\phi)/I_0\sim \sin(\pi+\phi)$ while the other is in a $0$-state revealing a CPR of the form $I(\phi)/I_0\sim \sin\phi$. 
Very intriguingly when the vertical bias is applied the first setup can pass a $0-\pi$ transition when $V_0\sim 1.5\Delta_0$ as we see in Fig. \ref{fig2}(a). However the second setup remains always in $0$-state and increasing the vertical bias only suppresses the current. Moreover we note that the amplitude of current in two cases is very different and for the considered parameters, setup (b) has almost three orders of magnitude larger supercurrent.
\begin{figure}[tp]
\includegraphics[width=0.85\linewidth]{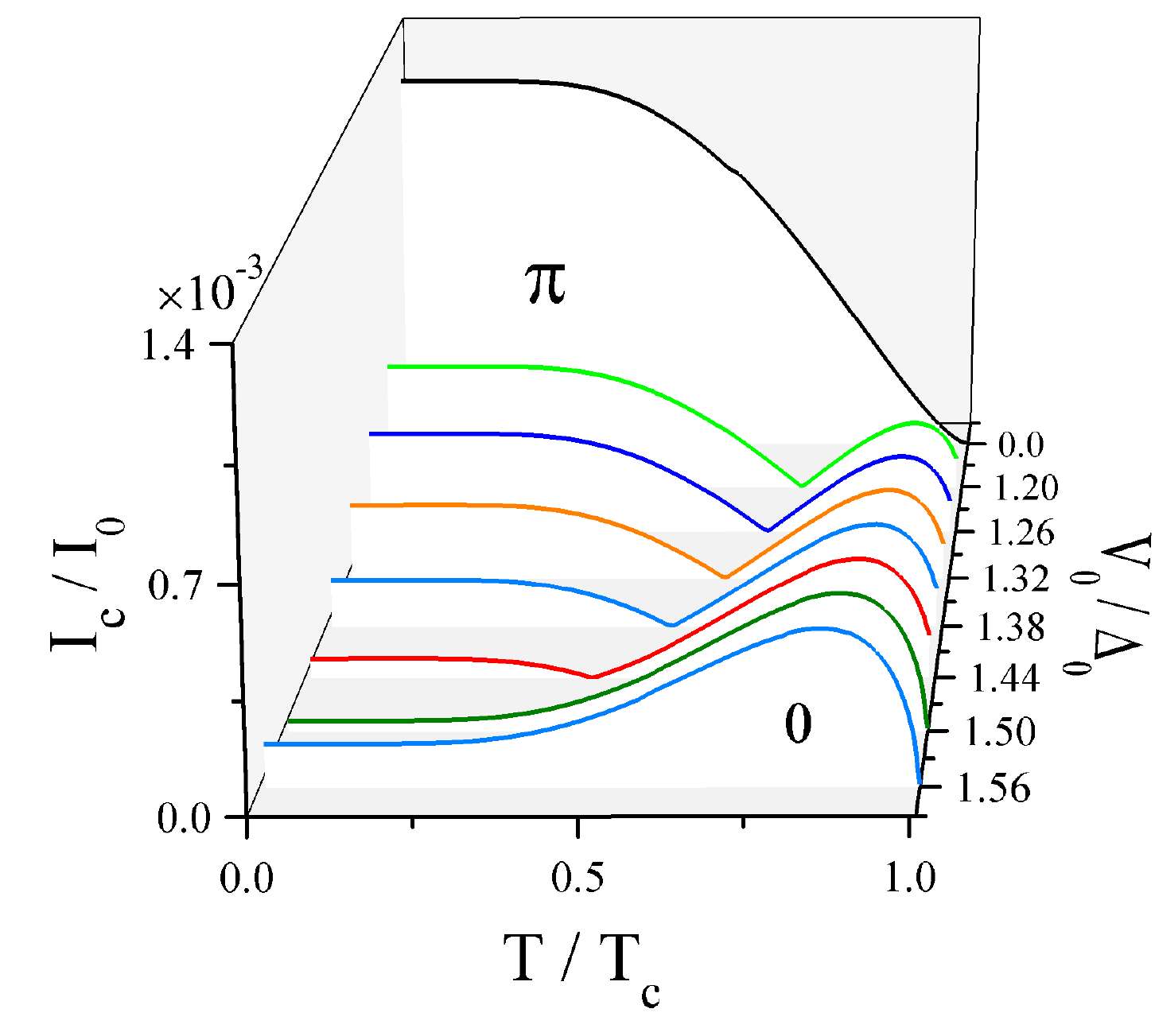}
\caption{
(Color online) Critical currents $I_{c}/I_{0}$ as function of normalized temperature $T/T_{c}$ are depicted versus different values of scaled bias voltage $V_{0}/t_{\perp}$, shifted on the $z$ axis, for the setup (a). The dimensionless length of the junction is fixed at value $d/l_{\perp}=4$. The Cusp in the curves represents the $\pi-0$ transition where the critical current vanish.}\label{fig3}
\end{figure} 
\par
It is worth to note that the main difference between two setups is the role of interlayer hopping $t_{\perp}$ in their transport properties even in normal (non-superconducting) state. In setup (a) $t_{\perp}$ plays a significant and constructive role by facilitating the pass way through the flake and one expect that if we could switch it off the current will vanish and transport through the BG regions is not possible. This could explain why the supercurrent in device (a) is much smaller than the device (b) since the electrons must change the layer and enter another transport channel. However, in the setup (b) the interlayer hopping acts only as a barrier for the movement of quasiparticles and it somehow provides a resistance against transmission through the scattering region. In fact in this case only the bottom ML is responsible for the transport and subsequently varying none of the parameters like vertical bias, the temperature, and doping leads to the qualitative change in the behavior of supercurrent and $0-\pi$ transitions cannot take place. In other words the upper layer of the flake only introduces an extra channel for scattering off the bottom layer which slightly suppresses the electron transport through the system. 
When $V_0$ is absent, the device (b) shows large transmission probabilities and as a result the CPR is strongly non-sinusoidal while increasing the vertical bias, which opens a gap in the band structure, leads to the decline in the supercurrent and CPR becomes closer to the sinusoidal behavior. 
\par
In the remainder we will concentrate on the first device to understand the underlying physics of supercurrent reversal and $0-\pi$ transitions which can occur by varying $V_0$ and $\mu$ of the BG flake as well as temperature and the junction length when either $V_0$ or $\mu$ have a finite value. We present the dependence of the critical current $I_c/I_0$ on the temperature varying the vertical bias scaled by the superconducting gap $V_0/\Delta_0$ in Fig. \ref{fig3}. 
When no vertical bias voltage is applied, the junction remains always in $\pi$ state for all temperatures below the critical temperature $T<T_c$. Nevertheless applying $V_0$ the supercurrent as the function of $T/T_c$ shows one cusp indicating a $\pi$ to $0$ transition. Moreover the position of the cusps varies with $V_0$ and after $V_0\sim 1.5\Delta_0$ the junction  will be completely in $0$ state irrespective of the temperature variations. As one can see from Fig. \ref{fig3} while at small values of vertical bias the critical current shows an overall decline, when it enters the $0$ state for a wide range of temperatures, $I_c$ can even increase with $T$. 
\par
In order to understand above mentioned features especially the mechanism of supercurrent reversal, we use the intuition based on the scattering matrix for the normal transport governed by ${\rm S}_B$ which contains the properties of the BG flake contacted by MG regions. We have already discussed that the transmission through the junction (a) is very small and subsequently the Andreev bound states are formed from single scatterings from the junction and the multiple scattering processes are strongly suppressed. Upon the electron hole conversion or vice versa at the NS interfaces the quasiparticles acquires an energy dependent Andreev phase  beside the superconducting phase $\pm\phi/2$. Moreover the excitations passing between two interfaces may find an extra phase shift $\gamma_{\rm sc}$, corresponding to the phase factors of the transmission coefficients $t'_{12}$ and $t_{21}$  defined by $t=|t|\exp(\gamma_{\rm sc})$. Therefore one can convince himself that the phase accumulated in the excitations during scattering $\gamma_{\rm sc}$ is added to the superconducting phase difference which leads to a shift in the $\phi$ dependence of the ABS energies. The occurrence of $\pi$ state for the device (a) when $\mu=V_0=0$ signals that there must be some phase shift of amount $\pi$ originating from $\gamma_{\rm sc}$ while in the other setup there is no phase shift. The phase $\gamma_{\rm sc}$ in the setup (a), in fact, originates from the transmission between lower and upper layers of the BG flake accompanied with the change of sublattice index $A_{1}\rightarrow B_{2}$. To understand its root let us have a look to the Hamiltonian (\ref{BG-H}) in which the two Dirac Blocks, corresponding to the two MLs constructing the bilayer, differs with each other. The difference comes from the fact that we have AB stacking and the upper layer is $\pi/3$ rotated with respect to the lower. Subsequently the wavefunctions spinor structures in two layers are not the same which leads to the emergence of an extra phase $\gamma_{\rm sc}$ when the electrons need to pass from lower to the upper layer. At this point one may wonder how the two setups become different since in both of them the excitations undergoes transitions between the layers several times depending on the length 
$d$. But as one can immediately see from Fig. \ref{fig1} in the second setup since at the end quasiparticles again leave the flake to the lower ML and we have the same number of passing from lower to the upper and vice versa there will be no net phase accumulated in the transmission coefficients. On the other hand in setup (a) there is always one more transition from lower to the upper layer and the scattered excitations from the junction will have a net phase factor. It must be mentioned that $\gamma_{\rm sc}$ in general depends on the angle of incidence of the particles, as well as the doping $\mu$ and bias $V_0$. Thus in the absence of the bias and doping the two setups are in $\pi$ and $0$ states, respectively, as the overall effect of scattering phases $\gamma_{\rm sc}$. It is worth to note that if we had AA stacked bilayer flake even in the first setup no phase factor appears due to the transition between the layers and subsequently the junction will be in the $0$ state.
\begin{figure}
\includegraphics[width=0.85\linewidth]{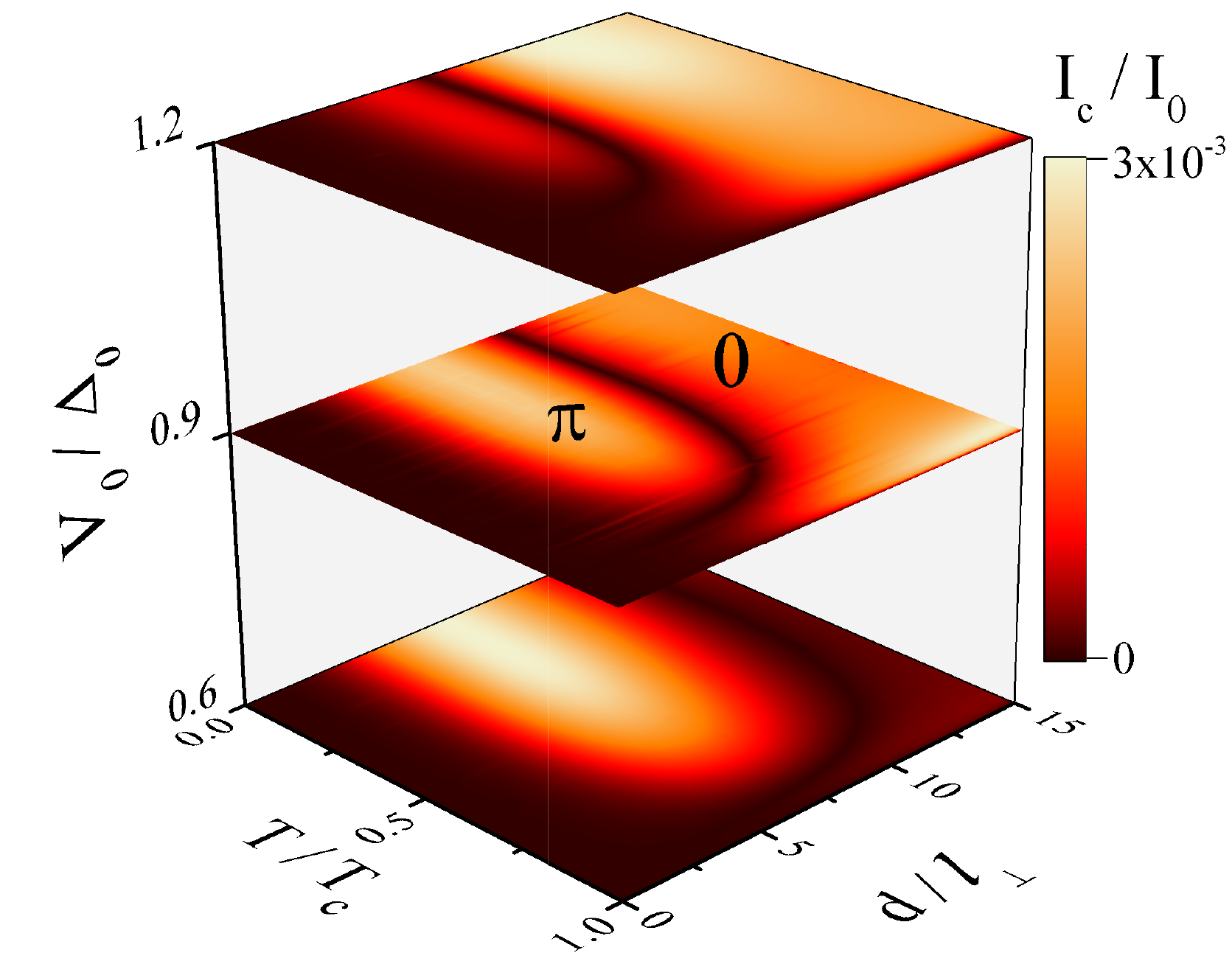}
\caption{(Color online) Critical currents $I_{c}/I_{0}$ versus
normalized temperature $T/T_{c}$ and the normalized length of the junction $d/l_{\perp}$ are plotted for the scaled bias values [0.6, 0.9, 1.2]. The Cusp in the curves represents the $\pi-0$ transition that again moves with the inclusion of topological term.}\label{fig4}
\end{figure} 
\par
We have already seen that when a vertical bias is applied, the junction undergoes a $\pi$ to $0$ transitions which is clear from both Figs. \ref{fig2} and \ref{fig3}. Up to now we have considered a junction with fixed length $d/l_\perp\sim 4$. But if we increase the length as shown in Fig. \ref{fig4} the Josephson current changes to a $0$ state in a certain length which depends on the strength of the vertical bias and the temperature. Moreover at the higher $V_0$ the domain of the $\pi$ junction versus $T/T_c$ and $d/l_\perp$ shrinks and becomes smaller. The bias induced transition originates from the dependence of the scattering phase $\gamma_{\rm sc}$ to $V_0$. In fact application of the bias between the layers leads to an asymmetry which influences the scattering processes and the phase of the different modes transmission coefficients such that after a certain value of the vertical bias depending on the length, the junction becomes of $0$ type. It must be noted that when the vertical bias voltage goes beyond the superconducting gap, the excitations carrying the supercurrent becomes evanescent due to the gap formation in the spectrum of BG flake. 
\par
\begin{figure}
\includegraphics[width=0.85\linewidth]{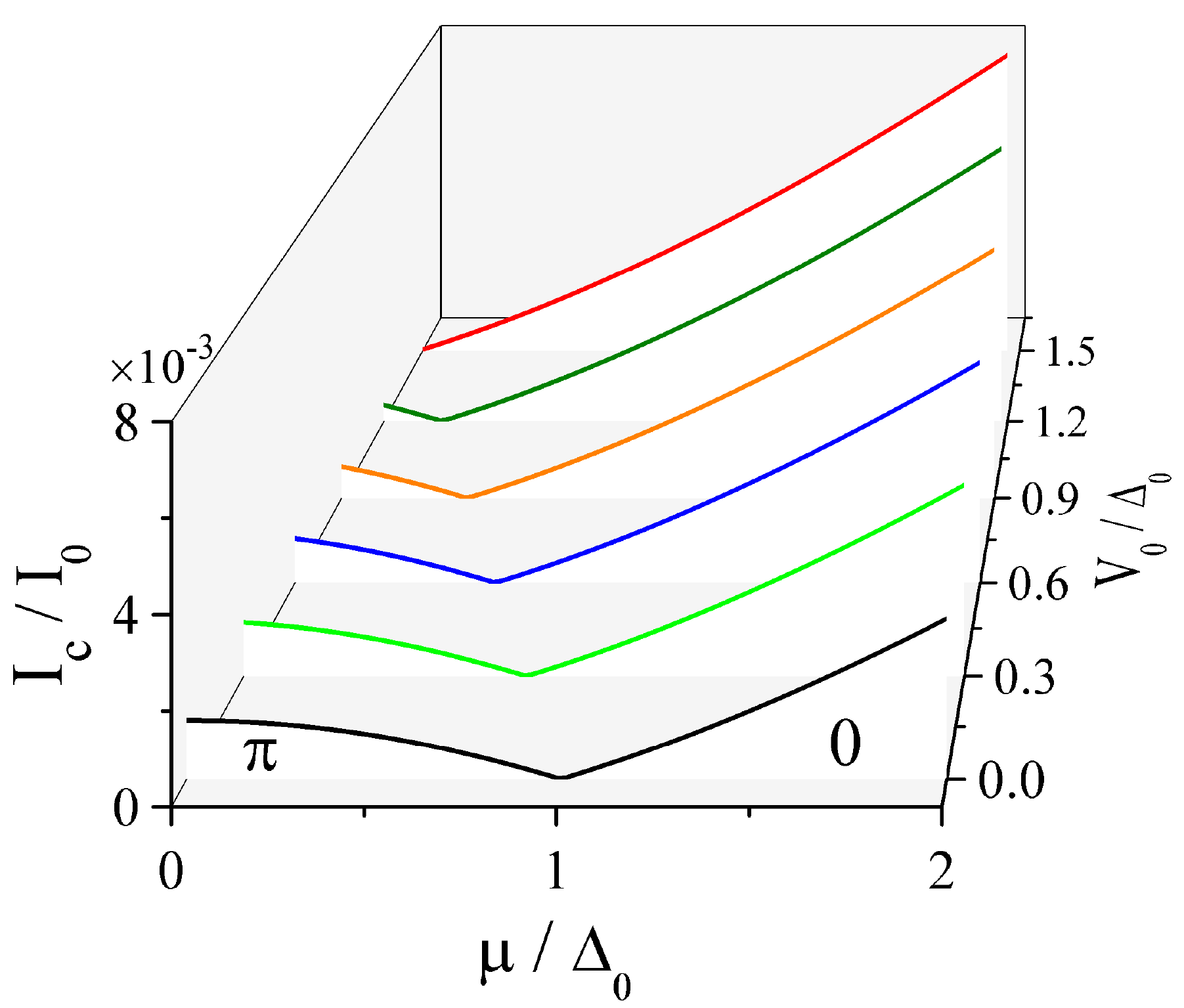}
\caption{(Color online) Critical currents $I_{c}/I_{0}$ as function of scaled doping of the Bg flake $\mu/\Delta_{0}$ are demonstrated versus different values of scaled bias voltage $V_{0}/t_{\perp}$, shifted on the $z$ axis, for the setup (a) at the zero temperature $T=0$. The dimensionless length of the junction is $d/l_{\perp}=4$. The Cusp in the curves indicates the $\pi-0$ transition.}\label{fig5}
\end{figure}

The effect of doping is somehow similar to the application of $V_0$ and leads to the $\pi-0$ transitions as indicated in Fig. \ref{fig5}. For small values of bias and doping in comparison with superconducting gap the junction remains in the $\pi$ state while when either $\mu$ or $V_0$ proceed well above $\Delta_0$ the supercurrent shows a $0$ behavior. In this state the critical current becomes larger upon increasing any of them. We can relate the transition with $\mu$ to the fact that by increasing the doping, more modes with different incidence angles participate effectively in transport and therefore their scattering phases $\gamma_{\rm sc}$ are washed out. This cause the junction to enter a $0$ state but due to the mechanism different from the bias induced $0$ state. 
\par
Finally we should comment on the possible experimental realization of the device under consideration and the predicted results. As it has been mentioned in the introduction, the step-like ML/BL interfaces and bilayer flakes connected to monolayer regions have been already observed and some transport experiments have been done based on them. On the other hand there are a large amount of experiments on the superconducting proximity effect and Josephson junctions based on graphene. So it seems that the device we propose here is completely reliable in the current experimental setups and the supercurrent reversal can be investigated in them by applying the vertical bias voltage, changing the doping and even the temperature.

\section{Conclusion}\label{sec:concl}
In this paper we employ scattering theory to study the supercurrent flowing through a weak link containing a biased BG flake embedded between two MG based superconducting reservoirs. We investigate two possible configurations for the setup regarding how the BG flake is attached to the MG regions in the left and right. The reservoirs are connected either to different layers of the BG flake  or to the same layer of the BG. The supercurrent passing through the undoped BG flake in the first situation is in the $\pi$ state while the other shows a $0$ behavior in the absence of the bias voltage ($V_{0}=0$). Both states are robust against varying the length of the junction $d$ as well as the temperature $T$ as long as $\mu=V_0=0$. We argue that the underlying mechanism of supercurrent reversal, characterized by the cusp in the critical current curves in the first setup, is related to the scattering phases accumulated in excitations upon transmission through the junction. We reveal that the biased BG flake in the second setup remains always in $0$ state and increasing the vertical bias only suppresses the current. Nevertheless, the biased junction in the first setup undergoes a $\pi-0$ transition at zero doping provided 
by a finite value of $V_{0}$ depending on the length. Increasing the doping causes $\pi-0$ transitions as well so that the junction is in $\pi$ state as long as $\mu$ and $V_0$ are small.

\acknowledgments
We are grateful to Boris Altshuler and Yuli Nazarov for useful discussions. BZR thanks Gerrit Bauer for the hospitality and support during his visit to the Kavli Institute of NanoScience, Delft University of Technology in Delft.

\bibliography{bma-stepgraph}

\begin{thebibliography}{55}%
\makeatletter
\providecommand \@ifxundefined [1]{%
 \@ifx{#1\undefined}
}%
\providecommand \@ifnum [1]{%
 \ifnum #1\expandafter \@firstoftwo
 \else \expandafter \@secondoftwo
 \fi
}%
\providecommand \@ifx [1]{%
 \ifx #1\expandafter \@firstoftwo
 \else \expandafter \@secondoftwo
 \fi
}%
\providecommand \natexlab [1]{#1}%
\providecommand \enquote  [1]{``#1''}%
\providecommand \bibnamefont  [1]{#1}%
\providecommand \bibfnamefont [1]{#1}%
\providecommand \citenamefont [1]{#1}%
\providecommand \href@noop [0]{\@secondoftwo}%
\providecommand \href [0]{\begingroup \@sanitize@url \@href}%
\providecommand \@href[1]{\@@startlink{#1}\@@href}%
\providecommand \@@href[1]{\endgroup#1\@@endlink}%
\providecommand \@sanitize@url [0]{\catcode `\\12\catcode `\$12\catcode
  `\&12\catcode `\#12\catcode `\^12\catcode `\_12\catcode `\%12\relax}%
\providecommand \@@startlink[1]{}%
\providecommand \@@endlink[0]{}%
\providecommand \url  [0]{\begingroup\@sanitize@url \@url }%
\providecommand \@url [1]{\endgroup\@href {#1}{\urlprefix }}%
\providecommand \urlprefix  [0]{URL }%
\providecommand \Eprint [0]{\href }%
\providecommand \doibase [0]{http://dx.doi.org/}%
\providecommand \selectlanguage [0]{\@gobble}%
\providecommand \bibinfo  [0]{\@secondoftwo}%
\providecommand \bibfield  [0]{\@secondoftwo}%
\providecommand \translation [1]{[#1]}%
\providecommand \BibitemOpen [0]{}%
\providecommand \bibitemStop [0]{}%
\providecommand \bibitemNoStop [0]{.\EOS\space}%
\providecommand \EOS [0]{\spacefactor3000\relax}%
\providecommand \BibitemShut  [1]{\csname bibitem#1\endcsname}%
\let\auto@bib@innerbib\@empty
\bibitem [{\citenamefont {Geim}\ and\ \citenamefont
  {Novoselov}(2007)}]{geim2007}%
  \BibitemOpen
  \bibfield  {author} {\bibinfo {author} {\bibfnamefont {A.~K.}\ \bibnamefont
  {Geim}}\ and\ \bibinfo {author} {\bibfnamefont {K.~S.}\ \bibnamefont
  {Novoselov}},\ }\href {\doibase 10.1038/nmat1849} {\bibfield  {journal}
  {\bibinfo  {journal} {Nature Mater.}\ }\textbf {\bibinfo {volume} {6}},\
  \bibinfo {pages} {183} (\bibinfo {year} {2007})}\BibitemShut {NoStop}%
\bibitem [{\citenamefont {Castro~Neto}\ \emph {et~al.}(2009)\citenamefont
  {Castro~Neto}, \citenamefont {Guinea}, \citenamefont {Peres}, \citenamefont
  {Novoselov},\ and\ \citenamefont {Geim}}]{neto2009}%
  \BibitemOpen
  \bibfield  {author} {\bibinfo {author} {\bibfnamefont {A.~H.}\ \bibnamefont
  {Castro~Neto}}, \bibinfo {author} {\bibfnamefont {F.}~\bibnamefont {Guinea}},
  \bibinfo {author} {\bibfnamefont {N.~M.~R.}\ \bibnamefont {Peres}}, \bibinfo
  {author} {\bibfnamefont {K.~S.}\ \bibnamefont {Novoselov}}, \ and\ \bibinfo
  {author} {\bibfnamefont {A.~K.}\ \bibnamefont {Geim}},\ }\href {\doibase
  10.1103/RevModPhys.81.109} {\bibfield  {journal} {\bibinfo  {journal} {Rev.
  Mod. Phys.}\ }\textbf {\bibinfo {volume} {81}},\ \bibinfo {pages} {109}
  (\bibinfo {year} {2009})}\BibitemShut {NoStop}%
\bibitem [{\citenamefont {Novoselov}\ \emph {et~al.}(2006)\citenamefont
  {Novoselov}, \citenamefont {McCann}, \citenamefont {Morozov}, \citenamefont
  {Fal'ko}, \citenamefont {Katsnelson}, \citenamefont {Zeitler}, \citenamefont
  {Schedin},\ and\ \citenamefont {Geim}}]{novoselov2006}%
  \BibitemOpen
  \bibfield  {author} {\bibinfo {author} {\bibfnamefont {K.~S.}\ \bibnamefont
  {Novoselov}}, \bibinfo {author} {\bibfnamefont {E.}~\bibnamefont {McCann}},
  \bibinfo {author} {\bibfnamefont {S.~V.}\ \bibnamefont {Morozov}}, \bibinfo
  {author} {\bibfnamefont {V.~I.}\ \bibnamefont {Fal'ko}}, \bibinfo {author}
  {\bibfnamefont {M.~I.}\ \bibnamefont {Katsnelson}}, \bibinfo {author}
  {\bibfnamefont {D.}~\bibnamefont {Zeitler}, \bibfnamefont {U.and~Jiang}},
  \bibinfo {author} {\bibfnamefont {F.}~\bibnamefont {Schedin}}, \ and\
  \bibinfo {author} {\bibfnamefont {A.~K.}\ \bibnamefont {Geim}},\ }\href
  {\doibase 10.1038/nphys245} {\bibfield  {journal} {\bibinfo  {journal} {Nat.
  Phys.}\ }\textbf {\bibinfo {volume} {2}},\ \bibinfo {pages} {177} (\bibinfo
  {year} {2006})}\BibitemShut {NoStop}%
\bibitem [{\citenamefont {McCann}\ \emph {et~al.}(2007)\citenamefont {McCann},
  \citenamefont {Abergel},\ and\ \citenamefont {Fal’ko}}]{mccann2007}%
  \BibitemOpen
  \bibfield  {author} {\bibinfo {author} {\bibfnamefont {E.}~\bibnamefont
  {McCann}}, \bibinfo {author} {\bibfnamefont {D.~S.}\ \bibnamefont {Abergel}},
  \ and\ \bibinfo {author} {\bibfnamefont {V.~I.}\ \bibnamefont {Fal’ko}},\
  }\href {\doibase http://dx.doi.org/10.1016/j.ssc.2007.03.054} {\bibfield
  {journal} {\bibinfo  {journal} {Solid State Commun.}\ }\textbf {\bibinfo
  {volume} {143}},\ \bibinfo {pages} {110 } (\bibinfo {year}
  {2007})}\BibitemShut {NoStop}%
\bibitem [{\citenamefont {McCann}\ and\ \citenamefont
  {Koshino}(2013)}]{mccann2013}%
  \BibitemOpen
  \bibfield  {author} {\bibinfo {author} {\bibfnamefont {E.}~\bibnamefont
  {McCann}}\ and\ \bibinfo {author} {\bibfnamefont {M.}~\bibnamefont
  {Koshino}},\ }\href {\doibase 10.1088/0034-4885/76/5/056503} {\bibfield
  {journal} {\bibinfo  {journal} {Rep. Prog. Phys.}\ }\textbf {\bibinfo
  {volume} {76}},\ \bibinfo {pages} {056503} (\bibinfo {year}
  {2013})}\BibitemShut {NoStop}%
\bibitem [{\citenamefont {Katsnelson}\ \emph {et~al.}(2006)\citenamefont
  {Katsnelson}, \citenamefont {Novoselov},\ and\ \citenamefont
  {Geim}}]{katsnelson2006}%
  \BibitemOpen
  \bibfield  {author} {\bibinfo {author} {\bibfnamefont {M.~I.}\ \bibnamefont
  {Katsnelson}}, \bibinfo {author} {\bibfnamefont {K.~S.}\ \bibnamefont
  {Novoselov}}, \ and\ \bibinfo {author} {\bibfnamefont {A.~K.}\ \bibnamefont
  {Geim}},\ }\href {\doibase 10.1038/nphys384} {\bibfield  {journal} {\bibinfo
  {journal} {Nat. Phys.}\ }\textbf {\bibinfo {volume} {2}},\ \bibinfo {pages}
  {620} (\bibinfo {year} {2006})}\BibitemShut {NoStop}%
\bibitem [{\citenamefont {Ohta}\ \emph {et~al.}(2006)\citenamefont {Ohta},
  \citenamefont {Bostwick}, \citenamefont {Seyller}, \citenamefont {Horn},\
  and\ \citenamefont {Rotenberg}}]{ohta2006}%
  \BibitemOpen
  \bibfield  {author} {\bibinfo {author} {\bibfnamefont {T.}~\bibnamefont
  {Ohta}}, \bibinfo {author} {\bibfnamefont {A.}~\bibnamefont {Bostwick}},
  \bibinfo {author} {\bibfnamefont {T.}~\bibnamefont {Seyller}}, \bibinfo
  {author} {\bibfnamefont {K.}~\bibnamefont {Horn}}, \ and\ \bibinfo {author}
  {\bibfnamefont {E.}~\bibnamefont {Rotenberg}},\ }\href {\doibase
  10.1126/science.1130681} {\bibfield  {journal} {\bibinfo  {journal}
  {Science}\ }\textbf {\bibinfo {volume} {313}},\ \bibinfo {pages} {951}
  (\bibinfo {year} {2006})}\BibitemShut {NoStop}%
\bibitem [{\citenamefont {Castro}\ \emph {et~al.}(2007)\citenamefont {Castro},
  \citenamefont {Novoselov}, \citenamefont {Morozov}, \citenamefont {Peres},
  \citenamefont {Lopes~dos Santos}, \citenamefont {Nilsson}, \citenamefont
  {Guinea}, \citenamefont {Geim},\ and\ \citenamefont
  {Castro~Neto}}]{castro2007}%
  \BibitemOpen
  \bibfield  {author} {\bibinfo {author} {\bibfnamefont {E.~V.}\ \bibnamefont
  {Castro}}, \bibinfo {author} {\bibfnamefont {K.~S.}\ \bibnamefont
  {Novoselov}}, \bibinfo {author} {\bibfnamefont {S.~V.}\ \bibnamefont
  {Morozov}}, \bibinfo {author} {\bibfnamefont {N.~M.~R.}\ \bibnamefont
  {Peres}}, \bibinfo {author} {\bibfnamefont {J.~M.~B.}\ \bibnamefont
  {Lopes~dos Santos}}, \bibinfo {author} {\bibfnamefont {J.}~\bibnamefont
  {Nilsson}}, \bibinfo {author} {\bibfnamefont {F.}~\bibnamefont {Guinea}},
  \bibinfo {author} {\bibfnamefont {A.~K.}\ \bibnamefont {Geim}}, \ and\
  \bibinfo {author} {\bibfnamefont {A.~H.}\ \bibnamefont {Castro~Neto}},\
  }\href {\doibase 10.1103/PhysRevLett.99.216802} {\bibfield  {journal}
  {\bibinfo  {journal} {Phys. Rev. Lett.}\ }\textbf {\bibinfo {volume} {99}},\
  \bibinfo {pages} {216802} (\bibinfo {year} {2007})}\BibitemShut {NoStop}%
\bibitem [{\citenamefont {Zhang}\ \emph {et~al.}(2009)\citenamefont {Zhang},
  \citenamefont {Tang}, \citenamefont {Girit}, \citenamefont {Hao},
  \citenamefont {Martin}, \citenamefont {Zettl}, \citenamefont {Crommie},
  \citenamefont {Shen},\ and\ \citenamefont {Wang}}]{wang2009-nat}%
  \BibitemOpen
  \bibfield  {author} {\bibinfo {author} {\bibfnamefont {Y.}~\bibnamefont
  {Zhang}}, \bibinfo {author} {\bibfnamefont {T.-T.}\ \bibnamefont {Tang}},
  \bibinfo {author} {\bibfnamefont {C.}~\bibnamefont {Girit}}, \bibinfo
  {author} {\bibfnamefont {Z.}~\bibnamefont {Hao}}, \bibinfo {author}
  {\bibfnamefont {M.~C.}\ \bibnamefont {Martin}}, \bibinfo {author}
  {\bibfnamefont {A.}~\bibnamefont {Zettl}}, \bibinfo {author} {\bibfnamefont
  {M.~F.}\ \bibnamefont {Crommie}}, \bibinfo {author} {\bibfnamefont {Y.~R.}\
  \bibnamefont {Shen}}, \ and\ \bibinfo {author} {\bibfnamefont
  {F.}~\bibnamefont {Wang}},\ }\href {\doibase 10.1038/nature08105} {\bibfield
  {journal} {\bibinfo  {journal} {Nature}\ }\textbf {\bibinfo {volume} {459}},\
  \bibinfo {pages} {820} (\bibinfo {year} {2009})}\BibitemShut {NoStop}%
\bibitem [{\citenamefont {Oostinga}\ \emph {et~al.}(2008)\citenamefont
  {Oostinga}, \citenamefont {Heersche}, \citenamefont {Liu}, \citenamefont
  {Morpurgo},\ and\ \citenamefont {Vandersypen}}]{oostinga2008}%
  \BibitemOpen
  \bibfield  {author} {\bibinfo {author} {\bibfnamefont {J.~B.}\ \bibnamefont
  {Oostinga}}, \bibinfo {author} {\bibfnamefont {H.~B.}\ \bibnamefont
  {Heersche}}, \bibinfo {author} {\bibfnamefont {X.}~\bibnamefont {Liu}},
  \bibinfo {author} {\bibfnamefont {A.~F.}\ \bibnamefont {Morpurgo}}, \ and\
  \bibinfo {author} {\bibfnamefont {L.~M.~K.}\ \bibnamefont {Vandersypen}},\
  }\href {\doibase 10.1038/nmat2082} {\bibfield  {journal} {\bibinfo  {journal}
  {Nat. Mater.}\ }\textbf {\bibinfo {volume} {7}},\ \bibinfo {pages} {151}
  (\bibinfo {year} {2008})}\BibitemShut {NoStop}%
\bibitem [{\citenamefont {McCann}(2006)}]{mccann2006}%
  \BibitemOpen
  \bibfield  {author} {\bibinfo {author} {\bibfnamefont {E.}~\bibnamefont
  {McCann}},\ }\href {\doibase 10.1103/PhysRevB.74.161403} {\bibfield
  {journal} {\bibinfo  {journal} {Phys. Rev. B}\ }\textbf {\bibinfo {volume}
  {74}},\ \bibinfo {pages} {161403} (\bibinfo {year} {2006})}\BibitemShut
  {NoStop}%
\bibitem [{\citenamefont {Min}\ \emph {et~al.}(2007)\citenamefont {Min},
  \citenamefont {Sahu}, \citenamefont {Banerjee},\ and\ \citenamefont
  {MacDonald}}]{min2007}%
  \BibitemOpen
  \bibfield  {author} {\bibinfo {author} {\bibfnamefont {H.}~\bibnamefont
  {Min}}, \bibinfo {author} {\bibfnamefont {B.}~\bibnamefont {Sahu}}, \bibinfo
  {author} {\bibfnamefont {S.~K.}\ \bibnamefont {Banerjee}}, \ and\ \bibinfo
  {author} {\bibfnamefont {A.~H.}\ \bibnamefont {MacDonald}},\ }\href {\doibase
  10.1103/PhysRevB.75.155115} {\bibfield  {journal} {\bibinfo  {journal} {Phys.
  Rev. B}\ }\textbf {\bibinfo {volume} {75}},\ \bibinfo {pages} {155115}
  (\bibinfo {year} {2007})}\BibitemShut {NoStop}%
\bibitem [{\citenamefont {Ji}\ \emph {et~al.}(2012)\citenamefont {Ji},
  \citenamefont {Hannon}, \citenamefont {Tromp}, \citenamefont {Perebeinos},
  \citenamefont {Tersoff},\ and\ \citenamefont {Ross}}]{ji2012}%
  \BibitemOpen
  \bibfield  {author} {\bibinfo {author} {\bibfnamefont {S.-H.}\ \bibnamefont
  {Ji}}, \bibinfo {author} {\bibfnamefont {J.~B.}\ \bibnamefont {Hannon}},
  \bibinfo {author} {\bibfnamefont {R.~M.}\ \bibnamefont {Tromp}}, \bibinfo
  {author} {\bibfnamefont {V.}~\bibnamefont {Perebeinos}}, \bibinfo {author}
  {\bibfnamefont {J.}~\bibnamefont {Tersoff}}, \ and\ \bibinfo {author}
  {\bibfnamefont {F.~M.}\ \bibnamefont {Ross}},\ }\href {\doibase
  10.1038/nmat3170} {\bibfield  {journal} {\bibinfo  {journal} {Nat Mater}\
  }\textbf {\bibinfo {volume} {11}},\ \bibinfo {pages} {114} (\bibinfo {year}
  {2012})}\BibitemShut {NoStop}%
\bibitem [{\citenamefont {Giannazzo}\ \emph {et~al.}(2012)\citenamefont
  {Giannazzo}, \citenamefont {Deretzis}, \citenamefont {La~Magna},
  \citenamefont {Roccaforte},\ and\ \citenamefont {Yakimova}}]{giannazzo2012}%
  \BibitemOpen
  \bibfield  {author} {\bibinfo {author} {\bibfnamefont {F.}~\bibnamefont
  {Giannazzo}}, \bibinfo {author} {\bibfnamefont {I.}~\bibnamefont {Deretzis}},
  \bibinfo {author} {\bibfnamefont {A.}~\bibnamefont {La~Magna}}, \bibinfo
  {author} {\bibfnamefont {F.}~\bibnamefont {Roccaforte}}, \ and\ \bibinfo
  {author} {\bibfnamefont {R.}~\bibnamefont {Yakimova}},\ }\href {\doibase
  10.1103/PhysRevB.86.235422} {\bibfield  {journal} {\bibinfo  {journal} {Phys.
  Rev. B}\ }\textbf {\bibinfo {volume} {86}},\ \bibinfo {pages} {235422}
  (\bibinfo {year} {2012})}\BibitemShut {NoStop}%
\bibitem [{\citenamefont {Clark}\ \emph {et~al.}(2014)\citenamefont {Clark},
  \citenamefont {Zhang}, \citenamefont {Gu}, \citenamefont {Park},
  \citenamefont {He}, \citenamefont {Feenstra},\ and\ \citenamefont
  {Li}}]{clark2014prx}%
  \BibitemOpen
  \bibfield  {author} {\bibinfo {author} {\bibfnamefont {K.~W.}\ \bibnamefont
  {Clark}}, \bibinfo {author} {\bibfnamefont {X.-G.}\ \bibnamefont {Zhang}},
  \bibinfo {author} {\bibfnamefont {G.}~\bibnamefont {Gu}}, \bibinfo {author}
  {\bibfnamefont {J.}~\bibnamefont {Park}}, \bibinfo {author} {\bibfnamefont
  {G.}~\bibnamefont {He}}, \bibinfo {author} {\bibfnamefont {R.~M.}\
  \bibnamefont {Feenstra}}, \ and\ \bibinfo {author} {\bibfnamefont {A.-P.}\
  \bibnamefont {Li}},\ }\href {\doibase 10.1103/PhysRevX.4.011021} {\bibfield
  {journal} {\bibinfo  {journal} {Phys. Rev. X}\ }\textbf {\bibinfo {volume}
  {4}},\ \bibinfo {pages} {011021} (\bibinfo {year} {2014})}\BibitemShut
  {NoStop}%
\bibitem [{\citenamefont {Nilsson}\ \emph {et~al.}(2007)\citenamefont
  {Nilsson}, \citenamefont {Castro~Neto}, \citenamefont {Guinea},\ and\
  \citenamefont {Peres}}]{nilsson2007}%
  \BibitemOpen
  \bibfield  {author} {\bibinfo {author} {\bibfnamefont {J.}~\bibnamefont
  {Nilsson}}, \bibinfo {author} {\bibfnamefont {A.~H.}\ \bibnamefont
  {Castro~Neto}}, \bibinfo {author} {\bibfnamefont {F.}~\bibnamefont {Guinea}},
  \ and\ \bibinfo {author} {\bibfnamefont {N.~M.~R.}\ \bibnamefont {Peres}},\
  }\href {\doibase 10.1103/PhysRevB.76.165416} {\bibfield  {journal} {\bibinfo
  {journal} {Phys. Rev. B}\ }\textbf {\bibinfo {volume} {76}},\ \bibinfo
  {pages} {165416} (\bibinfo {year} {2007})}\BibitemShut {NoStop}%
\bibitem [{\citenamefont {Nakanishi}\ \emph {et~al.}(2010)\citenamefont
  {Nakanishi}, \citenamefont {Koshino},\ and\ \citenamefont {Ando}}]{ando2010}%
  \BibitemOpen
  \bibfield  {author} {\bibinfo {author} {\bibfnamefont {T.}~\bibnamefont
  {Nakanishi}}, \bibinfo {author} {\bibfnamefont {M.}~\bibnamefont {Koshino}},
  \ and\ \bibinfo {author} {\bibfnamefont {T.}~\bibnamefont {Ando}},\ }\href
  {\doibase 10.1103/PhysRevB.82.125428} {\bibfield  {journal} {\bibinfo
  {journal} {Phys. Rev. B}\ }\textbf {\bibinfo {volume} {82}},\ \bibinfo
  {pages} {125428} (\bibinfo {year} {2010})}\BibitemShut {NoStop}%
\bibitem [{\citenamefont {Nakanishi}\ \emph {et~al.}(2011)\citenamefont
  {Nakanishi}, \citenamefont {Koshino},\ and\ \citenamefont {Ando}}]{ando2011}%
  \BibitemOpen
  \bibfield  {author} {\bibinfo {author} {\bibfnamefont {T.}~\bibnamefont
  {Nakanishi}}, \bibinfo {author} {\bibfnamefont {M.}~\bibnamefont {Koshino}},
  \ and\ \bibinfo {author} {\bibfnamefont {T.}~\bibnamefont {Ando}},\ }\href
  {\doibase 10.1088/1742-6596/302/1/012021} {\bibfield  {journal} {\bibinfo
  {journal} {J. Phys.: Conf. Ser.}\ }\textbf {\bibinfo {volume} {302}},\
  \bibinfo {pages} {012021} (\bibinfo {year} {2011})}\BibitemShut {NoStop}%
\bibitem [{\citenamefont {Koshino}\ \emph {et~al.}(2010)\citenamefont
  {Koshino}, \citenamefont {Nakanishi},\ and\ \citenamefont
  {Ando}}]{koshino2010}%
  \BibitemOpen
  \bibfield  {author} {\bibinfo {author} {\bibfnamefont {M.}~\bibnamefont
  {Koshino}}, \bibinfo {author} {\bibfnamefont {T.}~\bibnamefont {Nakanishi}},
  \ and\ \bibinfo {author} {\bibfnamefont {T.}~\bibnamefont {Ando}},\ }\href
  {\doibase 10.1103/PhysRevB.82.205436} {\bibfield  {journal} {\bibinfo
  {journal} {Phys. Rev. B}\ }\textbf {\bibinfo {volume} {82}},\ \bibinfo
  {pages} {205436} (\bibinfo {year} {2010})}\BibitemShut {NoStop}%
\bibitem [{\citenamefont {Tian}\ \emph {et~al.}(2013)\citenamefont {Tian},
  \citenamefont {Jiang}, \citenamefont {Childres}, \citenamefont {Cao},
  \citenamefont {Hu},\ and\ \citenamefont {Chen}}]{tian2013}%
  \BibitemOpen
  \bibfield  {author} {\bibinfo {author} {\bibfnamefont {J.}~\bibnamefont
  {Tian}}, \bibinfo {author} {\bibfnamefont {Y.}~\bibnamefont {Jiang}},
  \bibinfo {author} {\bibfnamefont {I.}~\bibnamefont {Childres}}, \bibinfo
  {author} {\bibfnamefont {H.}~\bibnamefont {Cao}}, \bibinfo {author}
  {\bibfnamefont {J.}~\bibnamefont {Hu}}, \ and\ \bibinfo {author}
  {\bibfnamefont {Y.~P.}\ \bibnamefont {Chen}},\ }\href {\doibase
  10.1103/PhysRevB.88.125410} {\bibfield  {journal} {\bibinfo  {journal} {Phys.
  Rev. B}\ }\textbf {\bibinfo {volume} {88}},\ \bibinfo {pages} {125410}
  (\bibinfo {year} {2013})}\BibitemShut {NoStop}%
\bibitem [{\citenamefont {Puls}\ \emph {et~al.}(2009)\citenamefont {Puls},
  \citenamefont {Staley},\ and\ \citenamefont {Liu}}]{puls2009}%
  \BibitemOpen
  \bibfield  {author} {\bibinfo {author} {\bibfnamefont {C.~P.}\ \bibnamefont
  {Puls}}, \bibinfo {author} {\bibfnamefont {N.~E.}\ \bibnamefont {Staley}}, \
  and\ \bibinfo {author} {\bibfnamefont {Y.}~\bibnamefont {Liu}},\ }\href
  {\doibase 10.1103/PhysRevB.79.235415} {\bibfield  {journal} {\bibinfo
  {journal} {Phys. Rev. B}\ }\textbf {\bibinfo {volume} {79}},\ \bibinfo
  {pages} {235415} (\bibinfo {year} {2009})}\BibitemShut {NoStop}%
\bibitem [{\citenamefont {Wei~Yan}\ \emph {et~al.}()\citenamefont {Wei~Yan},
  \citenamefont {Li}, \citenamefont {Yin}, \citenamefont {Qiao}, \citenamefont
  {Nie},\ and\ \citenamefont {He}}]{yan2015}%
  \BibitemOpen
  \bibfield  {author} {\bibinfo {author} {\bibfnamefont {W.}~\bibnamefont
  {Wei~Yan}}, \bibinfo {author} {\bibfnamefont {S.-Y.}\ \bibnamefont {Li}},
  \bibinfo {author} {\bibfnamefont {L.-J.}\ \bibnamefont {Yin}}, \bibinfo
  {author} {\bibfnamefont {J.-B.}\ \bibnamefont {Qiao}}, \bibinfo {author}
  {\bibfnamefont {J.-C.}\ \bibnamefont {Nie}}, \ and\ \bibinfo {author}
  {\bibfnamefont {L.}~\bibnamefont {He}},\ }\href@noop {} {\ }\BibitemShut
  {NoStop}%
\bibitem [{\citenamefont {Hasegawa}\ and\ \citenamefont
  {Kohmoto}(2012)}]{kohmoto2012}%
  \BibitemOpen
  \bibfield  {author} {\bibinfo {author} {\bibfnamefont {Y.}~\bibnamefont
  {Hasegawa}}\ and\ \bibinfo {author} {\bibfnamefont {M.}~\bibnamefont
  {Kohmoto}},\ }\href {\doibase 10.1103/PhysRevB.85.125430} {\bibfield
  {journal} {\bibinfo  {journal} {Phys. Rev. B}\ }\textbf {\bibinfo {volume}
  {85}},\ \bibinfo {pages} {125430} (\bibinfo {year} {2012})}\BibitemShut
  {NoStop}%
\bibitem [{\citenamefont {xiang Hu}\ and\ \citenamefont {Ding}(2012)}]{hu2012}%
  \BibitemOpen
  \bibfield  {author} {\bibinfo {author} {\bibfnamefont {Z.}~\bibnamefont
  {xiang Hu}}\ and\ \bibinfo {author} {\bibfnamefont {W.}~\bibnamefont
  {Ding}},\ }\href {\doibase http://dx.doi.org/10.1016/j.physleta.2011.11.046}
  {\bibfield  {journal} {\bibinfo  {journal} {Physics Letters A}\ }\textbf
  {\bibinfo {volume} {376}},\ \bibinfo {pages} {610 } (\bibinfo {year}
  {2012})}\BibitemShut {NoStop}%
\bibitem [{\citenamefont {Li}\ \emph {et~al.}(2011)\citenamefont {Li},
  \citenamefont {Li}, \citenamefont {Zheng},\ and\ \citenamefont
  {Niu}}]{li2011}%
  \BibitemOpen
  \bibfield  {author} {\bibinfo {author} {\bibfnamefont {H.}~\bibnamefont
  {Li}}, \bibinfo {author} {\bibfnamefont {H.}~\bibnamefont {Li}}, \bibinfo
  {author} {\bibfnamefont {Y.}~\bibnamefont {Zheng}}, \ and\ \bibinfo {author}
  {\bibfnamefont {J.}~\bibnamefont {Niu}},\ }\href {\doibase
  http://dx.doi.org/10.1016/j.physb.2011.01.034} {\bibfield  {journal}
  {\bibinfo  {journal} {Physica B: Condensed Matter}\ }\textbf {\bibinfo
  {volume} {406}},\ \bibinfo {pages} {1385 } (\bibinfo {year}
  {2011})}\BibitemShut {NoStop}%
\bibitem [{\citenamefont {Yin}\ \emph {et~al.}(2013)\citenamefont {Yin},
  \citenamefont {Liu}, \citenamefont {Li}, \citenamefont {Geng}, \citenamefont
  {Wang},\ and\ \citenamefont {Huai}}]{yin2013}%
  \BibitemOpen
  \bibfield  {author} {\bibinfo {author} {\bibfnamefont {D.}~\bibnamefont
  {Yin}}, \bibinfo {author} {\bibfnamefont {W.}~\bibnamefont {Liu}}, \bibinfo
  {author} {\bibfnamefont {X.}~\bibnamefont {Li}}, \bibinfo {author}
  {\bibfnamefont {L.}~\bibnamefont {Geng}}, \bibinfo {author} {\bibfnamefont
  {X.}~\bibnamefont {Wang}}, \ and\ \bibinfo {author} {\bibfnamefont
  {P.}~\bibnamefont {Huai}},\ }\href {\doibase
  http://dx.doi.org/10.1063/1.4826694} {\bibfield  {journal} {\bibinfo
  {journal} {Applied Physics Letters}\ }\textbf {\bibinfo {volume} {103}},\
  \bibinfo {eid} {173519} (\bibinfo {year} {2013})}\BibitemShut {NoStop}%
\bibitem [{\citenamefont {Wang}(2014)}]{wang2014}%
  \BibitemOpen
  \bibfield  {author} {\bibinfo {author} {\bibfnamefont {Y.}~\bibnamefont
  {Wang}},\ }\href {\doibase http://dx.doi.org/10.1063/1.4900731} {\bibfield
  {journal} {\bibinfo  {journal} {Journal of Applied Physics}\ }\textbf
  {\bibinfo {volume} {116}},\ \bibinfo {eid} {164317} (\bibinfo {year}
  {2014})}\BibitemShut {NoStop}%
\bibitem [{\citenamefont {Berahman}\ \emph {et~al.}(2014)\citenamefont
  {Berahman}, \citenamefont {Sanaee},\ and\ \citenamefont
  {Ghayour}}]{berahman2014}%
  \BibitemOpen
  \bibfield  {author} {\bibinfo {author} {\bibfnamefont {M.}~\bibnamefont
  {Berahman}}, \bibinfo {author} {\bibfnamefont {M.}~\bibnamefont {Sanaee}}, \
  and\ \bibinfo {author} {\bibfnamefont {R.}~\bibnamefont {Ghayour}},\ }\href
  {\doibase http://dx.doi.org/10.1016/j.carbon.2014.04.020} {\bibfield
  {journal} {\bibinfo  {journal} {Carbon}\ }\textbf {\bibinfo {volume} {75}},\
  \bibinfo {pages} {411 } (\bibinfo {year} {2014})}\BibitemShut {NoStop}%
\bibitem [{\citenamefont {Dragoman}(2013)}]{dragoman2013}%
  \BibitemOpen
  \bibfield  {author} {\bibinfo {author} {\bibfnamefont {D.}~\bibnamefont
  {Dragoman}},\ }\href {\doibase http://dx.doi.org/10.1063/1.4808904}
  {\bibfield  {journal} {\bibinfo  {journal} {Journal of Applied Physics}\
  }\textbf {\bibinfo {volume} {113}},\ \bibinfo {eid} {214312} (\bibinfo {year}
  {2013})}\BibitemShut {NoStop}%
\bibitem [{\citenamefont {Gonz\'alez}\ \emph {et~al.}(2010)\citenamefont
  {Gonz\'alez}, \citenamefont {Santos}, \citenamefont {Pacheco}, \citenamefont
  {Chico},\ and\ \citenamefont {Brey}}]{brey2010}%
  \BibitemOpen
  \bibfield  {author} {\bibinfo {author} {\bibfnamefont {J.~W.}\ \bibnamefont
  {Gonz\'alez}}, \bibinfo {author} {\bibfnamefont {H.}~\bibnamefont {Santos}},
  \bibinfo {author} {\bibfnamefont {M.}~\bibnamefont {Pacheco}}, \bibinfo
  {author} {\bibfnamefont {L.}~\bibnamefont {Chico}}, \ and\ \bibinfo {author}
  {\bibfnamefont {L.}~\bibnamefont {Brey}},\ }\href {\doibase
  10.1103/PhysRevB.81.195406} {\bibfield  {journal} {\bibinfo  {journal} {Phys.
  Rev. B}\ }\textbf {\bibinfo {volume} {81}},\ \bibinfo {pages} {195406}
  (\bibinfo {year} {2010})}\BibitemShut {NoStop}%
\bibitem [{\citenamefont {Beenakker}(2006)}]{beenakker2006}%
  \BibitemOpen
  \bibfield  {author} {\bibinfo {author} {\bibfnamefont {C.~W.~J.}\
  \bibnamefont {Beenakker}},\ }\href {\doibase 10.1103/PhysRevLett.97.067007}
  {\bibfield  {journal} {\bibinfo  {journal} {Phys. Rev. Lett.}\ }\textbf
  {\bibinfo {volume} {97}},\ \bibinfo {pages} {067007} (\bibinfo {year}
  {2006})}\BibitemShut {NoStop}%
\bibitem [{\citenamefont {Beenakker}(2008)}]{beenakker2008}%
  \BibitemOpen
  \bibfield  {author} {\bibinfo {author} {\bibfnamefont {C.~W.~J.}\
  \bibnamefont {Beenakker}},\ }\href {\doibase 10.1103/RevModPhys.80.1337}
  {\bibfield  {journal} {\bibinfo  {journal} {Rev. Mod. Phys.}\ }\textbf
  {\bibinfo {volume} {80}},\ \bibinfo {pages} {1337} (\bibinfo {year}
  {2008})}\BibitemShut {NoStop}%
\bibitem [{\citenamefont {Heersche}\ \emph {et~al.}(2007)\citenamefont
  {Heersche}, \citenamefont {Jarillo-Herrero}, \citenamefont {Oostinga},
  \citenamefont {Vandersypen},\ and\ \citenamefont {Morpurgo}}]{heersche}%
  \BibitemOpen
  \bibfield  {author} {\bibinfo {author} {\bibfnamefont {H.~B.}\ \bibnamefont
  {Heersche}}, \bibinfo {author} {\bibfnamefont {P.}~\bibnamefont
  {Jarillo-Herrero}}, \bibinfo {author} {\bibfnamefont {J.~B.}\ \bibnamefont
  {Oostinga}}, \bibinfo {author} {\bibfnamefont {L.~M.~K.}\ \bibnamefont
  {Vandersypen}}, \ and\ \bibinfo {author} {\bibfnamefont {A.~F.}\ \bibnamefont
  {Morpurgo}},\ }\href {\doibase 10.1038/nature05555} {\bibfield  {journal}
  {\bibinfo  {journal} {Nature}\ }\textbf {\bibinfo {volume} {446}},\ \bibinfo
  {pages} {56} (\bibinfo {year} {2007})}\BibitemShut {NoStop}%
\bibitem [{\citenamefont {Titov}\ and\ \citenamefont
  {Beenakker}(2006)}]{titov2006}%
  \BibitemOpen
  \bibfield  {author} {\bibinfo {author} {\bibfnamefont {M.}~\bibnamefont
  {Titov}}\ and\ \bibinfo {author} {\bibfnamefont {C.~W.~J.}\ \bibnamefont
  {Beenakker}},\ }\href {\doibase 10.1103/PhysRevB.74.041401} {\bibfield
  {journal} {\bibinfo  {journal} {Phys. Rev. B}\ }\textbf {\bibinfo {volume}
  {74}},\ \bibinfo {pages} {041401} (\bibinfo {year} {2006})}\BibitemShut
  {NoStop}%
\bibitem [{\citenamefont {Moghaddam}\ and\ \citenamefont
  {Zareyan}(2006)}]{moghaddam2006}%
  \BibitemOpen
  \bibfield  {author} {\bibinfo {author} {\bibfnamefont {A.~G.}\ \bibnamefont
  {Moghaddam}}\ and\ \bibinfo {author} {\bibfnamefont {M.}~\bibnamefont
  {Zareyan}},\ }\href {\doibase 10.1103/PhysRevB.74.241403} {\bibfield
  {journal} {\bibinfo  {journal} {Phys. Rev. B}\ }\textbf {\bibinfo {volume}
  {74}},\ \bibinfo {pages} {241403} (\bibinfo {year} {2006})}\BibitemShut
  {NoStop}%
\bibitem [{\citenamefont {Rainis}\ \emph {et~al.}(2009)\citenamefont {Rainis},
  \citenamefont {Taddei}, \citenamefont {Dolcini}, \citenamefont {Polini},\
  and\ \citenamefont {Fazio}}]{fazio2009}%
  \BibitemOpen
  \bibfield  {author} {\bibinfo {author} {\bibfnamefont {D.}~\bibnamefont
  {Rainis}}, \bibinfo {author} {\bibfnamefont {F.}~\bibnamefont {Taddei}},
  \bibinfo {author} {\bibfnamefont {F.}~\bibnamefont {Dolcini}}, \bibinfo
  {author} {\bibfnamefont {M.}~\bibnamefont {Polini}}, \ and\ \bibinfo {author}
  {\bibfnamefont {R.}~\bibnamefont {Fazio}},\ }\href {\doibase
  10.1103/PhysRevB.79.115131} {\bibfield  {journal} {\bibinfo  {journal} {Phys.
  Rev. B}\ }\textbf {\bibinfo {volume} {79}},\ \bibinfo {pages} {115131}
  (\bibinfo {year} {2009})}\BibitemShut {NoStop}%
\bibitem [{\citenamefont {Linder}\ and\ \citenamefont
  {Sudb\o{}}(2007)}]{linder2007}%
  \BibitemOpen
  \bibfield  {author} {\bibinfo {author} {\bibfnamefont {J.}~\bibnamefont
  {Linder}}\ and\ \bibinfo {author} {\bibfnamefont {A.}~\bibnamefont
  {Sudb\o{}}},\ }\href {\doibase 10.1103/PhysRevLett.99.147001} {\bibfield
  {journal} {\bibinfo  {journal} {Phys. Rev. Lett.}\ }\textbf {\bibinfo
  {volume} {99}},\ \bibinfo {pages} {147001} (\bibinfo {year}
  {2007})}\BibitemShut {NoStop}%
\bibitem [{\citenamefont {Linder}\ \emph {et~al.}(2009)\citenamefont {Linder},
  \citenamefont {Black-Schaffer}, \citenamefont {Yokoyama}, \citenamefont
  {Doniach},\ and\ \citenamefont {Sudb\o{}}}]{linder2009}%
  \BibitemOpen
  \bibfield  {author} {\bibinfo {author} {\bibfnamefont {J.}~\bibnamefont
  {Linder}}, \bibinfo {author} {\bibfnamefont {A.~M.}\ \bibnamefont
  {Black-Schaffer}}, \bibinfo {author} {\bibfnamefont {T.}~\bibnamefont
  {Yokoyama}}, \bibinfo {author} {\bibfnamefont {S.}~\bibnamefont {Doniach}}, \
  and\ \bibinfo {author} {\bibfnamefont {A.}~\bibnamefont {Sudb\o{}}},\ }\href
  {\doibase 10.1103/PhysRevB.80.094522} {\bibfield  {journal} {\bibinfo
  {journal} {Phys. Rev. B}\ }\textbf {\bibinfo {volume} {80}},\ \bibinfo
  {pages} {094522} (\bibinfo {year} {2009})}\BibitemShut {NoStop}%
\bibitem [{\citenamefont {Black-Schaffer}\ and\ \citenamefont
  {Doniach}(2008)}]{doniach2008}%
  \BibitemOpen
  \bibfield  {author} {\bibinfo {author} {\bibfnamefont {A.~M.}\ \bibnamefont
  {Black-Schaffer}}\ and\ \bibinfo {author} {\bibfnamefont {S.}~\bibnamefont
  {Doniach}},\ }\href {\doibase 10.1103/PhysRevB.78.024504} {\bibfield
  {journal} {\bibinfo  {journal} {Phys. Rev. B}\ }\textbf {\bibinfo {volume}
  {78}},\ \bibinfo {pages} {024504} (\bibinfo {year} {2008})}\BibitemShut
  {NoStop}%
\bibitem [{\citenamefont {Herrera}\ \emph {et~al.}(2010)\citenamefont
  {Herrera}, \citenamefont {Burset},\ and\ \citenamefont
  {Yeyati}}]{yeyati2010}%
  \BibitemOpen
  \bibfield  {author} {\bibinfo {author} {\bibfnamefont {W.~J.}\ \bibnamefont
  {Herrera}}, \bibinfo {author} {\bibfnamefont {P.}~\bibnamefont {Burset}}, \
  and\ \bibinfo {author} {\bibfnamefont {A.~L.}\ \bibnamefont {Yeyati}},\
  }\href {\doibase 10.1088/0953-8984/22/27/275304} {\bibfield  {journal}
  {\bibinfo  {journal} {Journal of Physics: Condensed Matter}\ }\textbf
  {\bibinfo {volume} {22}},\ \bibinfo {pages} {275304} (\bibinfo {year}
  {2010})}\BibitemShut {NoStop}%
\bibitem [{\citenamefont {Ojeda-Aristizabal}\ \emph {et~al.}(2009)\citenamefont
  {Ojeda-Aristizabal}, \citenamefont {Ferrier}, \citenamefont {Gu\'eron},\ and\
  \citenamefont {Bouchiat}}]{bouchiat2009}%
  \BibitemOpen
  \bibfield  {author} {\bibinfo {author} {\bibfnamefont {C.}~\bibnamefont
  {Ojeda-Aristizabal}}, \bibinfo {author} {\bibfnamefont {M.}~\bibnamefont
  {Ferrier}}, \bibinfo {author} {\bibfnamefont {S.}~\bibnamefont {Gu\'eron}}, \
  and\ \bibinfo {author} {\bibfnamefont {H.}~\bibnamefont {Bouchiat}},\ }\href
  {\doibase 10.1103/PhysRevB.79.165436} {\bibfield  {journal} {\bibinfo
  {journal} {Phys. Rev. B}\ }\textbf {\bibinfo {volume} {79}},\ \bibinfo
  {pages} {165436} (\bibinfo {year} {2009})}\BibitemShut {NoStop}%
\bibitem [{\citenamefont {Girit}\ \emph {et~al.}(2009)\citenamefont {Girit},
  \citenamefont {Bouchiat}, \citenamefont {Naaman}, \citenamefont {Zhang},
  \citenamefont {Crommie}, \citenamefont {Zettl},\ and\ \citenamefont
  {Siddiqi}}]{girit2009}%
  \BibitemOpen
  \bibfield  {author} {\bibinfo {author} {\bibfnamefont {c.}~\bibnamefont
  {Girit}}, \bibinfo {author} {\bibfnamefont {V.}~\bibnamefont {Bouchiat}},
  \bibinfo {author} {\bibfnamefont {O.}~\bibnamefont {Naaman}}, \bibinfo
  {author} {\bibfnamefont {Y.}~\bibnamefont {Zhang}}, \bibinfo {author}
  {\bibfnamefont {M.~F.}\ \bibnamefont {Crommie}}, \bibinfo {author}
  {\bibfnamefont {A.}~\bibnamefont {Zettl}}, \ and\ \bibinfo {author}
  {\bibfnamefont {I.}~\bibnamefont {Siddiqi}},\ }\href {\doibase
  10.1021/nl802765x} {\bibfield  {journal} {\bibinfo  {journal} {Nano Letters}\
  }\textbf {\bibinfo {volume} {9}},\ \bibinfo {pages} {198} (\bibinfo {year}
  {2009})}\BibitemShut {NoStop}%
\bibitem [{\citenamefont {Dirks}\ \emph {et~al.}(2011)\citenamefont {Dirks},
  \citenamefont {Hughes}, \citenamefont {Lal}, \citenamefont {Uchoa},
  \citenamefont {Chen}, \citenamefont {Chialvo}, \citenamefont {Goldbart},\
  and\ \citenamefont {Mason}}]{dirks2011}%
  \BibitemOpen
  \bibfield  {author} {\bibinfo {author} {\bibfnamefont {T.}~\bibnamefont
  {Dirks}}, \bibinfo {author} {\bibfnamefont {T.~L.}\ \bibnamefont {Hughes}},
  \bibinfo {author} {\bibfnamefont {S.}~\bibnamefont {Lal}}, \bibinfo {author}
  {\bibfnamefont {B.}~\bibnamefont {Uchoa}}, \bibinfo {author} {\bibfnamefont
  {Y.-F.}\ \bibnamefont {Chen}}, \bibinfo {author} {\bibfnamefont
  {C.}~\bibnamefont {Chialvo}}, \bibinfo {author} {\bibfnamefont {P.~M.}\
  \bibnamefont {Goldbart}}, \ and\ \bibinfo {author} {\bibfnamefont
  {N.}~\bibnamefont {Mason}},\ }\href {\doibase 10.1038/nphys1911} {\bibfield
  {journal} {\bibinfo  {journal} {Nat. Phys.}\ }\textbf {\bibinfo {volume} {7}}
  (\bibinfo {year} {2011}),\ 10.1038/nphys1911}\BibitemShut {NoStop}%
\bibitem [{\citenamefont {Jeong}\ \emph {et~al.}(2011)\citenamefont {Jeong},
  \citenamefont {Choi}, \citenamefont {Lee}, \citenamefont {Jo}, \citenamefont
  {Doh},\ and\ \citenamefont {Lee}}]{jeong2011}%
  \BibitemOpen
  \bibfield  {author} {\bibinfo {author} {\bibfnamefont {D.}~\bibnamefont
  {Jeong}}, \bibinfo {author} {\bibfnamefont {J.-H.}\ \bibnamefont {Choi}},
  \bibinfo {author} {\bibfnamefont {G.-H.}\ \bibnamefont {Lee}}, \bibinfo
  {author} {\bibfnamefont {S.}~\bibnamefont {Jo}}, \bibinfo {author}
  {\bibfnamefont {Y.-J.}\ \bibnamefont {Doh}}, \ and\ \bibinfo {author}
  {\bibfnamefont {H.-J.}\ \bibnamefont {Lee}},\ }\href {\doibase
  10.1103/PhysRevB.83.094503} {\bibfield  {journal} {\bibinfo  {journal} {Phys.
  Rev. B}\ }\textbf {\bibinfo {volume} {83}},\ \bibinfo {pages} {094503}
  (\bibinfo {year} {2011})}\BibitemShut {NoStop}%
\bibitem [{\citenamefont {Lee}\ \emph {et~al.}(2011)\citenamefont {Lee},
  \citenamefont {Jeong}, \citenamefont {Choi}, \citenamefont {Doh},\ and\
  \citenamefont {Lee}}]{lee2011}%
  \BibitemOpen
  \bibfield  {author} {\bibinfo {author} {\bibfnamefont {G.-H.}\ \bibnamefont
  {Lee}}, \bibinfo {author} {\bibfnamefont {D.}~\bibnamefont {Jeong}}, \bibinfo
  {author} {\bibfnamefont {J.-H.}\ \bibnamefont {Choi}}, \bibinfo {author}
  {\bibfnamefont {Y.-J.}\ \bibnamefont {Doh}}, \ and\ \bibinfo {author}
  {\bibfnamefont {H.-J.}\ \bibnamefont {Lee}},\ }\href {\doibase
  10.1103/PhysRevLett.107.146605} {\bibfield  {journal} {\bibinfo  {journal}
  {Phys. Rev. Lett.}\ }\textbf {\bibinfo {volume} {107}},\ \bibinfo {pages}
  {146605} (\bibinfo {year} {2011})}\BibitemShut {NoStop}%
\bibitem [{\citenamefont {Lee}\ \emph {et~al.}(2015)\citenamefont {Lee},
  \citenamefont {Kim}, \citenamefont {Jhi},\ and\ \citenamefont
  {Lee}}]{lee2015}%
  \BibitemOpen
  \bibfield  {author} {\bibinfo {author} {\bibfnamefont {G.-H.}\ \bibnamefont
  {Lee}}, \bibinfo {author} {\bibfnamefont {S.}~\bibnamefont {Kim}}, \bibinfo
  {author} {\bibfnamefont {S.-H.}\ \bibnamefont {Jhi}}, \ and\ \bibinfo
  {author} {\bibfnamefont {H.-J.}\ \bibnamefont {Lee}},\ }\href {\doibase
  10.1038/ncomms7181} {\bibfield  {journal} {\bibinfo  {journal} {Nat.
  Commun.}\ }\textbf {\bibinfo {volume} {6}},\ \bibinfo {pages} {6181}
  (\bibinfo {year} {2015})}\BibitemShut {NoStop}%
\bibitem [{\citenamefont {Zhang}\ \emph {et~al.}(2008)\citenamefont {Zhang},
  \citenamefont {Fu}, \citenamefont {Wang}, \citenamefont {Zhang},\ and\
  \citenamefont {Xing}}]{xing2008}%
  \BibitemOpen
  \bibfield  {author} {\bibinfo {author} {\bibfnamefont {Q.}~\bibnamefont
  {Zhang}}, \bibinfo {author} {\bibfnamefont {D.}~\bibnamefont {Fu}}, \bibinfo
  {author} {\bibfnamefont {B.}~\bibnamefont {Wang}}, \bibinfo {author}
  {\bibfnamefont {R.}~\bibnamefont {Zhang}}, \ and\ \bibinfo {author}
  {\bibfnamefont {D.~Y.}\ \bibnamefont {Xing}},\ }\href {\doibase
  10.1103/PhysRevLett.101.047005} {\bibfield  {journal} {\bibinfo  {journal}
  {Phys. Rev. Lett.}\ }\textbf {\bibinfo {volume} {101}},\ \bibinfo {pages}
  {047005} (\bibinfo {year} {2008})}\BibitemShut {NoStop}%
\bibitem [{\citenamefont {Bai}\ \emph {et~al.}(2008)\citenamefont {Bai},
  \citenamefont {Yang},\ and\ \citenamefont {Zhang}}]{bai2008}%
  \BibitemOpen
  \bibfield  {author} {\bibinfo {author} {\bibfnamefont {C.}~\bibnamefont
  {Bai}}, \bibinfo {author} {\bibfnamefont {Y.}~\bibnamefont {Yang}}, \ and\
  \bibinfo {author} {\bibfnamefont {X.}~\bibnamefont {Zhang}},\ }\href
  {\doibase http://dx.doi.org/10.1063/1.2894513} {\bibfield  {journal}
  {\bibinfo  {journal} {Applied Physics Letters}\ }\textbf {\bibinfo {volume}
  {92}},\ \bibinfo {eid} {102513} (\bibinfo {year} {2008})}\BibitemShut
  {NoStop}%
\bibitem [{\citenamefont {Cheng}\ \emph {et~al.}(2009)\citenamefont {Cheng},
  \citenamefont {Xing}, \citenamefont {Wang},\ and\ \citenamefont
  {Sun}}]{sun2009}%
  \BibitemOpen
  \bibfield  {author} {\bibinfo {author} {\bibfnamefont {S.-g.}\ \bibnamefont
  {Cheng}}, \bibinfo {author} {\bibfnamefont {Y.}~\bibnamefont {Xing}},
  \bibinfo {author} {\bibfnamefont {J.}~\bibnamefont {Wang}}, \ and\ \bibinfo
  {author} {\bibfnamefont {Q.-f.}\ \bibnamefont {Sun}},\ }\href {\doibase
  10.1103/PhysRevLett.103.167003} {\bibfield  {journal} {\bibinfo  {journal}
  {Phys. Rev. Lett.}\ }\textbf {\bibinfo {volume} {103}},\ \bibinfo {pages}
  {167003} (\bibinfo {year} {2009})}\BibitemShut {NoStop}%
\bibitem [{\citenamefont {Schelter}\ \emph {et~al.}(2012)\citenamefont
  {Schelter}, \citenamefont {Trauzettel},\ and\ \citenamefont
  {Recher}}]{recher2012}%
  \BibitemOpen
  \bibfield  {author} {\bibinfo {author} {\bibfnamefont {J.}~\bibnamefont
  {Schelter}}, \bibinfo {author} {\bibfnamefont {B.}~\bibnamefont
  {Trauzettel}}, \ and\ \bibinfo {author} {\bibfnamefont {P.}~\bibnamefont
  {Recher}},\ }\href {\doibase 10.1103/PhysRevLett.108.106603} {\bibfield
  {journal} {\bibinfo  {journal} {Phys. Rev. Lett.}\ }\textbf {\bibinfo
  {volume} {108}},\ \bibinfo {pages} {106603} (\bibinfo {year}
  {2012})}\BibitemShut {NoStop}%
\bibitem [{\citenamefont {Linder}\ \emph {et~al.}(2008)\citenamefont {Linder},
  \citenamefont {Yokoyama}, \citenamefont {Huertas-Hernando},\ and\
  \citenamefont {Sudb\o{}}}]{linder2008}%
  \BibitemOpen
  \bibfield  {author} {\bibinfo {author} {\bibfnamefont {J.}~\bibnamefont
  {Linder}}, \bibinfo {author} {\bibfnamefont {T.}~\bibnamefont {Yokoyama}},
  \bibinfo {author} {\bibfnamefont {D.}~\bibnamefont {Huertas-Hernando}}, \
  and\ \bibinfo {author} {\bibfnamefont {A.}~\bibnamefont {Sudb\o{}}},\ }\href
  {\doibase 10.1103/PhysRevLett.100.187004} {\bibfield  {journal} {\bibinfo
  {journal} {Phys. Rev. Lett.}\ }\textbf {\bibinfo {volume} {100}},\ \bibinfo
  {pages} {187004} (\bibinfo {year} {2008})}\BibitemShut {NoStop}%
\bibitem [{\citenamefont {Moghaddam}\ and\ \citenamefont
  {Zareyan}(2008)}]{moghaddam2008}%
  \BibitemOpen
  \bibfield  {author} {\bibinfo {author} {\bibfnamefont {A.~G.}\ \bibnamefont
  {Moghaddam}}\ and\ \bibinfo {author} {\bibfnamefont {M.}~\bibnamefont
  {Zareyan}},\ }\href {\doibase 10.1103/PhysRevB.78.115413} {\bibfield
  {journal} {\bibinfo  {journal} {Phys. Rev. B}\ }\textbf {\bibinfo {volume}
  {78}},\ \bibinfo {pages} {115413} (\bibinfo {year} {2008})}\BibitemShut
  {NoStop}%
\bibitem [{\citenamefont {Asano}\ \emph {et~al.}(2008)\citenamefont {Asano},
  \citenamefont {Yoshida}, \citenamefont {Tanaka},\ and\ \citenamefont
  {Golubov}}]{asano2008}%
  \BibitemOpen
  \bibfield  {author} {\bibinfo {author} {\bibfnamefont {Y.}~\bibnamefont
  {Asano}}, \bibinfo {author} {\bibfnamefont {T.}~\bibnamefont {Yoshida}},
  \bibinfo {author} {\bibfnamefont {Y.}~\bibnamefont {Tanaka}}, \ and\ \bibinfo
  {author} {\bibfnamefont {A.~A.}\ \bibnamefont {Golubov}},\ }\href {\doibase
  10.1103/PhysRevB.78.014514} {\bibfield  {journal} {\bibinfo  {journal} {Phys.
  Rev. B}\ }\textbf {\bibinfo {volume} {78}},\ \bibinfo {pages} {014514}
  (\bibinfo {year} {2008})}\BibitemShut {NoStop}%
\bibitem [{\citenamefont {Moghaddam}\ and\ \citenamefont
  {Zareyan}(2009)}]{moghaddam2009}%
  \BibitemOpen
  \bibfield  {author} {\bibinfo {author} {\bibfnamefont {A.~G.}\ \bibnamefont
  {Moghaddam}}\ and\ \bibinfo {author} {\bibfnamefont {M.}~\bibnamefont
  {Zareyan}},\ }\href {\doibase http://dx.doi.org/10.1016/j.ssc.2009.02.048}
  {\bibfield  {journal} {\bibinfo  {journal} {Solid State Commun.}\ }\textbf
  {\bibinfo {volume} {149}},\ \bibinfo {pages} {1106 } (\bibinfo {year}
  {2009})}\BibitemShut {NoStop}%
\bibitem [{\citenamefont {Rameshti}\ \emph {et~al.}(2014)\citenamefont
  {Rameshti}, \citenamefont {Moghaddam},\ and\ \citenamefont
  {Zareyan}}]{zare2014}%
  \BibitemOpen
  \bibfield  {author} {\bibinfo {author} {\bibfnamefont {B.~Z.}\ \bibnamefont
  {Rameshti}}, \bibinfo {author} {\bibfnamefont {A.~G.}\ \bibnamefont
  {Moghaddam}}, \ and\ \bibinfo {author} {\bibfnamefont {M.}~\bibnamefont
  {Zareyan}},\ }\href {\doibase 10.1209/0295-5075/108/37002} {\bibfield
  {journal} {\bibinfo  {journal} {Europhys. Lett.}\ }\textbf {\bibinfo {volume}
  {108}},\ \bibinfo {pages} {37002} (\bibinfo {year} {2014})}\BibitemShut
  {NoStop}%
\end{thebibliography}%

\end{document}